\documentclass[twocolumn,useAMS,usenatbib]{mn2e}
\usepackage{graphicx,amssymb,amsmath}
\usepackage{bm}

\newcommand{\rmd}{{\rm d}}
\newcommand{\rme}{{\rm e}}
\newcommand{\rmi}{{\rm i}}

\newcommand{\opL}{{I\!\!L}}

\newcommand{\cmnt}[1]{}

\title[Lindblad resonances II]
{Lindblad resonance torques in relativistic discs: II. Computation of resonance strengths}

\author[Hirata]
 {Christopher M. Hirata
\\Caltech M/C 350-17, Pasadena, California 91125, USA}

\date{22 February 2011}

\begin{document}
\maketitle

\begin{abstract}
We present a fully relativistic computation of the torques due to Lindblad resonances from perturbers on circular, equatorial orbits on discs around Schwarzschild and Kerr black holes.  The computation proceeds by establishing a relation between the Lindblad torques and the gravitational waveforms emitted by the perturber and a test particle in a slightly eccentric orbit at the radius of the Lindblad resonance.  We show that our result reduces to the usual formula when taking the nonrelativistic limit.  Discs around a black hole possess an $m=1$ inner Lindblad resonance (ILR) with no Newtonian Keplerian analogue; however its strength is very weak even in the moderately relativistic regime ($r/M\sim$ few tens), which is in part due to the partial cancellation of the two leading contributions to the resonant amplitude (the gravitoelectric octupole and gravitomagnetic quadrupole).  For equatorial orbits around Kerr black holes, we find that the $m=1$ ILR strength is enhanced for retrograde spins and suppressed for prograde spins.  We also find that the torque associated with the $m\ge 2$ ILRs is enhanced relative to the nonrelativistic case; the enhancement is a factor of 2 for the Schwarzschild hole even when the perturber is at a radius of $25M$.
\end{abstract}


\section{Introduction}
\label{sec:intro}

This is the second in a series of two papers devoted to a relativistic computation of torques from an external perturber on a thin disc due to interactions at the Lindblad resonances, i.e. locations in the disc where the orbital frequency $\Omega$ and the radial epicyclic frequency $\kappa$ satisfy $\kappa = \pm m(\Omega-\Omega_{\rm s})$, where $\Omega_{\rm s}$ is the pattern speed of the perturbation.  Such resonances have been extensively studied in the nonrelativistic case \citep[e.g.][]{1972MNRAS.157....1L, 1978ApJ...222..850G, 1979ApJ...233..857G, 1980ApJ...241..425G, 1979MNRAS.186..799L}.  In the first paper (``Paper I''), we performed this computation for a general time-stationary, axisymmetric, spacetime with an equatorial plane of symmetry and a metric perturbation $h_{\alpha\beta}$ that respects the equatorial symmetry.  This paper (``Paper II'') completes the evaluation of the Lindblad torque in the case of most interest: the perturbation of the accretion disc surrounding a Schwarzschild or Kerr black hole by a small secondary also orbiting in the equatorial plane.  Such computations of the Lindblad resonant strengths may be relevant in the context of electromagnetic counterparts to binary black hole mergers, particularly if an inner disc is involved \citep{2009arXiv0906.0825C}.  (The more complicated case of perturbations outside of the equatorial plane -- as may occur in the case of a merger where the primary hole is rotating and the secondary is in an inclined orbit -- is left to future work.)

The resonant torque formula in Paper I depended on the geodesic properties in the unperturbed spacetime as well as being proportional to the square of the absolute value of the resonant amplitude ${\cal S}^{(m)}$, which was a function of the $\rme^{\rmi m\phi}$ Fourier component of the metric perturbation $h_{\alpha\beta}$ and its spatial derivative $h_{\alpha\beta,r}$.  The construction of these perturbations generally depends on the solution for the Weyl tensor component $\psi_4$, which may be solved using a separable wave equation with a source given by the stress-energy tensor associated with the perturber \citep{1973ApJ...185..635T}; and then $h_{\alpha\beta}$ may be obtained by applying a second-order differential operator to a master potential \citep{1975PhRvD..11.2042C}, which may be derived from $\psi_4$ \citep{1978PhRvL..41..203W}.  Fortunately, for our computations there is a way to circumvent the \citet{1975PhRvD..11.2042C} procedure: Paper I showed that the particular combination of metric perturbations we require is related to ${\cal P}^{(m)}$, the power delivered to a test particle in a slightly eccentric orbit by the $\rme^{\rmi m\phi}$ component of the perturbation.  By replacing the perturber with an equivalent gravitational wave source -- either incoming from past null infinity in the case of an inner Lindblad resonance (ILR), or emerging from the past horizon in the case of an outer Lindblad resonance (OLR) -- we may equate ${\cal P}^{(m)}$ with the power absorbed from the gravitational wave.  However, energy is conserved on a time-independent background metric, and thus ${\cal P}^{(m)}$ can be related to the interference between the equivalent gravitational wave representing the perturbation and the gravitational wave emitted by the test particle.  This allows us to express the resonant amplitude and hence the resonant torque in terms of the waveforms emitted by the perturber and the test particle (both to future null infinity and into the future horizon), so that standard methods to solve for $\psi_4$ are sufficient.

The outline of this paper is as follows.  In Section~\ref{sec:metric}, we introduce the Kerr metric and review the associated standard notation.  Section~\ref{sec:geodesics} reviews the geodesics in the Kerr spacetime and their description with action-angle variables, and Section~\ref{sec:modes} describes the compuation of the perturbation in the Weyl scalar $\psi_4$; while both of these subjects are standard, there are some differences in our treatment that are particularly suited to the problem at hand, and we make frequent use of intermediate results when taking the nonrelativistic limit, so an extended discussion is warranted.  Section~\ref{sec:ResonantAmp} presents the key new theoretical result of this paper, relating the behaviour of $\psi_4$ near the horizon and at infinity to the resonant amplitude ${\cal S}^{(m)}$.  We recompute the resonant amplitudes in the Kepler problem in Section~\ref{sec:Kepler}, and then proceed to investigate the Lindblad resonances in the Schwarzschild problem in Section~\ref{sec:Schwarzschild}.  Section~\ref{sec:Kerr} then considers the Lindblad resonance amplitudes associated with equatorial orbits in the Kerr spacetime.  We conclude in Section~\ref{sec:discussion}.

\section{Kerr metric and notation}
\label{sec:metric}

\subsection{The metric and null tetrad}

We parameterize the Kerr black hole sequence with the gravitational mass $M$ and the specific angular momentum $a$.  We use relativistic units where the Newtonian gravitational constant and the speed of light are equal to unity.  The dimensionless angular momentum is $a_\star\equiv a/M$.

The Kerr metric in Boyer-Lindquist coordinates \citep{1967JMP.....8..265B} is
\begin{eqnarray}
ds^2 &=& 
-\left( 1 - \frac{2Mr}\Sigma \right) \rmd t^2 - \frac{4Mar}\Sigma\sin^2\theta\, \rmd t\rmd\phi
\nonumber \\ &&
 + \frac{(r^2+a^2)^2
-\Delta a^2\sin^2\theta}\Sigma\,\sin^2\theta\,\rmd\phi^2
\nonumber \\ && + \frac\Sigma\Delta\,\rmd r^2 + \Sigma\,\rmd\theta^2,
\end{eqnarray}
where $\Delta \equiv r^2 - 2Mr + a^2$ and
$\Sigma \equiv r^2+a^2\cos^2\theta$.
The contravariant metric coefficients are
\begin{eqnarray}
g^{tt} \!\!\!\! &=& \!\!\!\! -\frac{(r^2+a^2)^2-\Delta a^2\sin^2\theta}{\Delta\Sigma},
\nonumber \\
g^{t\phi} \!\!\!\! &=& \!\!\!\! -\frac{2aMr}{\Delta\Sigma},
{\rm ~~~~}
g^{\phi\phi} = \frac{\Delta-a^2\sin^2\theta}{\Delta\Sigma\sin^2\theta},
\nonumber \\
g^{rr} \!\!\!\! &=& \!\!\!\! \frac\Delta\Sigma,
{\rm~~and~~}
g^{\theta\theta} = \Sigma^{-1}.
\end{eqnarray}

The standard Newman-Penrose basis is
\begin{eqnarray}
{\bmath l} &=& \frac{r^2+a^2}\Delta \partial_t + \frac a\Delta \partial_\phi + \partial_r,
\nonumber \\
{\bmath n} &=& \frac\Delta{2\Sigma}\left( \frac{r^2+a^2}\Delta \partial_t + \frac a\Delta \partial_\phi - \partial_r\right),
\nonumber \\
{\bmath m} &=& \frac{\rmi a \sin\theta\,\partial_t + \partial_\theta + \rmi\csc\theta \,\partial_\phi}{\sqrt2\,(r+\rmi a\cos\theta)},
{\rm \,\,and}\nonumber \\
\bar{\bmath m} &=& \frac{-\rmi a \sin\theta\,\partial_t + \partial_\theta - \rmi\csc\theta \,\partial_\phi}{\sqrt2\,(r-\rmi a\cos\theta)}.
\end{eqnarray}
Expressions involving ${\bmath m}$ can be simplified if we use
\begin{equation}
\rho = \frac{-1}{r-\rmi a\cos\theta} {\rm ~~and~~}
\bar\rho = \frac{-1}{r+\rmi a\cos\theta},
\end{equation}
which satisfy $\rho\bar\rho=\Sigma^{-1}$.  The Weyl scalar $\psi_4$ used to describe the emitted gravitational waveform is
\begin{equation}
\psi_4 = -C_{\alpha\beta\gamma\delta} n^\alpha \bar m^\beta n^\gamma \bar m^\delta
= - R_{\alpha\beta\gamma\delta} n^\alpha \bar m^\beta n^\gamma \bar m^\delta,
\end{equation}
where $C_{\alpha\beta\gamma\delta}$ is the Weyl tensor, and the equivalence to the component formed from the Riemann tensor $R_{\alpha\beta\gamma\delta}$ is due to the Newman-Penrose basis conditions.

The horizons of the black hole are at radial coordinate
\begin{equation}
r_{{\rm h}\pm} = M \pm \sqrt{M^2-a^2}.
\end{equation}
Particles very close to the horizon ($r-r_{{\rm h}+}\rightarrow 0^+$) rotate at a pattern speed of the hole's angular velocity:
\begin{equation}
\Omega_{\rm H} = \frac{a}{2Mr_{{\rm h}+}} = \frac{a}{r_{{\rm h}+}^2+a^2} = \frac{1-\sqrt{1-a_\star^2}}{2a_\star M}.
\end{equation}

Note that for real coordinates, $\bar{\bmath m}={\bmath m}^\ast$ and $\bar\rho=\rho^\ast$, where $^\ast$ denotes the complex conjugate; however we will occasionally analytically continue $r$ to complex values, in which case the barred quantities are not the complex conjugates of the unbarred quantities: $\bar\rho(r,\theta)=\rho^\ast(r^\ast,\theta^\ast)\neq\rho^\ast(r,\theta)$.

Finally, we define
\begin{equation}
K\equiv\omega(r^2+a^2) - am
\end{equation}
and use the angular operator
\begin{equation}
{\opL}^\dagger_n \equiv \partial_\theta - m\csc\theta+a\omega\sin\theta+n\cot\theta.
\end{equation}

\subsection{Notation in related works}

Our notation appears to be common in the literature but other examples can be found.
\begin{itemize}
\item We are consistent with the metric and (where applicable) null tetrad used in the standard general relativity text by \citet{1984ucp..book.....W}.  \citet{1973grav.book.....M} use ``$\rho^2$'' to denote our $\Sigma$, and do not fix a normalization for the principal null vectors.
\item \citet{1992mtbh.book.....C} uses the $+---$ signature, and uses ``$\rho^2$'' to denote our $\Sigma$; ``$\varphi$'' to denote our $\phi$; ``$\Sigma^2$'' to denote our $(r^2+a^2)^2-\Delta a^2\sin^2\theta$; and ``$\delta$'' to denote our $\sin^2\theta$.  For the perturbations, \citet{1992mtbh.book.....C} denotes the frequency by $-\sigma^{\!_+}$, and uses the opposite sign of $K$.  Additionally, our $\rho$ and $\bar\rho$ are denoted by $-\bar\rho^{-1\,\ast}$ and $-\bar\rho^{-1}$, respectively.  However, the null tetrad and the operators $\opL_n$ and $\opL_n^\dagger$ are the same.
\end{itemize}

\section{Timelike geodesics in Kerr}
\label{sec:geodesics}

We utilize the Hamiltonian formulation of the equations of motion for a particle.  As is well-known, the action for a particle of mass $\mu$ is $S=-\mu\int d\tau$, where $\tau$ is the proper time along the particle trajectory.  For our purposes, the fastest route to the torque formula is {\em not} to use the covariant representation of the action but rather to explicitly parameterize the particle's trajectory using the coordinate time $t$, which is always possible outside the outer horizon.  This method, which explicitly keeps only the 3 physical degrees of freedom, is best suited to a perturbation analysis.

As in Newtonian perturbation theory analyses, it is most convenient to work with action-angle variables, using the 3+1 version of the Hamiltonian that retains no gauge freedom associated with the particle trajectory.  \citet{2008PhRvD..78f4028H} constructed a set of action-angle variables in which the particle's trajectory is parameterized by {\em proper} time $\tau$, and $t$ is promoted to a dynamical variable (with conjugate momentum $p_t=-\mu{\cal E}$).  Their actions $(J_r,J_\theta,J_\phi)$ are equal to ours, since the momenta are the same, however the angle variables are different since ours advance at uniform rate with respect to coordinate time and theirs advance at a uniform rate with respect to proper time.  Thus the Fourier decompositions are also different.  Other works that have constructed the Hamiltonian for geodesic motion in 4-dimensional space have projected the motion into the 3 physical degrees of freedom \citep{2002CQGra..19.2743S}, but appear not to have constructed the full transformation from action-angle variables to the familiar spatial coordinates and momenta, which we will need to complete here.  \citet{2010arXiv1009.4923F} considered resonances in inspiralling black hole binaries, but parameterize their trajectory in terms of the ``Mino time'' $\lambda = \int \rmd\tau/\Sigma$ \citep{2003PhRvD..67h4027M, 2005CQGra..22S.801D}.  This again means that they have an additional conjugate variable pair not present in our treatment, and that their angle variables advance at a constant rate as measured by $\lambda$ rather than by $t$.

\subsection{Hamiltonian and constants of the motion}

The trajectory of a massive particle can be followed by parameterizing the trajectory $x^i(t)$ where $x^i\in\{r,\theta,\phi\}$, and using the action $S=-\tau$.  This results in the Hamiltonian $H=-p_t$, where $p_t$ is determined from the $p_i$ via the mass-shell condition (Paper I):
\begin{equation}
H(t,x^i,p_i) = \frac{g^{ti}p_i - \sqrt{(g^{ti}p_i)^2 - g^{tt}g^{ij}p_ip_j - \mu^2g^{tt}}}{g^{tt}}.
\label{eq:H-explicit}
\end{equation}

The timelike geodesics in the Kerr metric are characterized by three constants: the energy per unit mass ${\cal E} = -p_t/\mu = -u_t$; the angular momentum around the symmetry axis per unit mass, ${\cal L} = p_\phi/\mu = u_\phi$; and the Carter constant,
\begin{equation}
{\cal Q} = 2\Sigma({\bmath u}\cdot {\bmath l})({\bmath u}\cdot{\bmath n}) - r^2 - ({\cal L}-a{\cal E})^2,
\end{equation}
which may also be expressed using
\begin{equation}
{\cal K} = {\cal Q} + ({\cal L}-a{\cal E})^2.
\end{equation}

Given the three constants of the motion $\{{\cal E},{\cal Q},{\cal L}\}$, it is possible to obtain the momenta when the particle passes through any spatial position $(r,\theta,\phi)$.  Specifically, we always have eastward momentum $u_\phi={\cal L}$.  The southward momentum given by Eq.~(7.164) of \citet{1992mtbh.book.....C},
\begin{equation}
u_\theta^2 = {\cal Q} - a^2(1-{\cal E}^2)\cos^2\theta - {\cal L}^2\cot^2\theta,
\label{eq:u-theta}
\end{equation}
and the radial momentum by Eq.~(7.160) of \citet{1992mtbh.book.....C},
\begin{equation}
\Delta^2 u_r^2 = [(r^2+a^2){\cal E}-a{\cal L}]^2 - \Delta(r^2 + {\cal K}).
\label{eq:u-r}
\end{equation}

\subsection{Actions in terms of the energy, Carter constant, and angular momentum}

It is useful in integrable problems to define the action-angle variables.  We begin by considering the actions corresponding to the $r$, $\theta$, and $\phi$ loops around the invariant torus corresponding to a set of constants $\{{\cal E},{\cal Q},{\cal L}\}$.  The $\phi$-direction is the easiest: the action is
\begin{equation}
J_\phi = \frac1{2\pi} \oint_0^{2\pi} p_\phi \,\rmd\phi = p_\phi = \mu{\cal L}.
\end{equation}
We define the notation $\tilde J_\phi\equiv J_\phi/\mu = {\cal L}$.

For the $\theta$-direction, we use Eq.~(\ref{eq:u-theta}), which defines a loop in the $(\theta,u_\theta)$-plane.  Its area,
\begin{equation}
\tilde J_\theta = \frac1{2\pi}\oint u_\theta\,\rmd\theta,
\label{eq:J-theta}
\end{equation}
involves an elliptic function, which however is most easily evaluated by numerical integration.  It is convenient to switch to the variable $z=\cos\theta$, in which case we find
\begin{equation}
(1-z^2)u_\theta^2 = {\cal Q}(1-z^2) - a^2(1-{\cal E}^2)z^2(1-z^2) - {\cal L}^2z^2.
\end{equation}
The turning points are found at the zeroes of the right-hand side, which is quadratic in $z^2$.  These zeroes are $z^2=z^2_\pm$; inspection of the sign of the right-hand side at $z^2\in\{0,1,\infty\}$ shows that the zeroes have the ordering $0<z_-^2<1<z_+^2$.  These zeroes can then be found by bisection.


We may then change variables from $\theta$ to $z$; noting that $u_z=(1-z^2)^{-1/2}u_\theta$, we find
\begin{equation}
u_z^2 = {\cal Q}(1-z^2)^{-2}\left( 1 - \frac{z^2}{z_-^2} \right) \left( 1 - \frac{z^2}{z_+^2} \right).
\label{eq:uz2}
\end{equation}
The action is then
\begin{equation}
\tilde J_\theta = \frac{{\cal Q}^{1/2}}{2\pi} \oint \sqrt{\left( 1 - \frac{z^2}{z_-^2} \right) \left( 1 - \frac{z^2}{z_+^2} \right)}\,\frac{\rmd z}{1-z^2}.
\end{equation}
We solve this integral with the substitution $z=z_-\sin(\frac12\pi\tanh\xi)$, where the full integral is given by 4 times the integral $\int_0^\infty \rmd\xi$.  Written in terms of $\xi$, the integrand is smooth, even, and decays exponentially at large $\xi$.  Summation of the integrand in $\xi$ at points $(n+\frac12)\Delta\xi$ thus enables evaluation of the integral with exponentially small error as $\Delta\xi\rightarrow 0^+$ and $N\Delta\xi\rightarrow\infty$ (where $N$ is the number of points).

For the $r$-direction, Eq.~(\ref{eq:u-r}) defines a loop in the $(r,u_r)$-plane, and one may again find the area
\begin{equation}
\tilde J_r = \frac1{2\pi}\oint u_r\,\rmd r.
\label{eq:J-r}
\end{equation}
A practical solution for $\tilde J_r$ is to find the turning points $r_-$ and $r_+$ by solving the quartic equation $\Delta^2u_r^2=0$ for $r$, Eq.~(\ref{eq:u-r}).\footnote{We solve the equation by first finding the inflection points (via a quadratic equation) and then using the bisection method to find the extrema.  Finally a further bisection gives the roots.  The sign pattern of the extrema determines whether there are 1 or 3 roots outside the outer horizon; stable bound orbits require 3 roots.}  Then a substitution of the form
\begin{equation}
r = \frac{r_++r_-}2 - \frac{r_+-r_-}2\tanh \beta
\label{eq:r-beta}
\end{equation}
enables one to turn the integral into one over $-\infty<\beta<\infty$ (multiplied by 2 to get the inward leg of the trajectory), where the integrand is analytic in the vicinity of the real $\beta$-axis and declines exponentially as $\beta\rightarrow\pm\infty$; it may thus be evaluated by the simple method of summing the integrand at equally spaced abscissae $\beta$.

A problem one may encounter is that there is only a finite range of energies $[{\cal E}_{\rm min}({\cal L},Q), {\cal E}_{\rm max}({\cal L},Q)]$ over which bound orbits can exist.  The sign pattern of the extrema can be used to distinguish the ${\cal E}<{\cal E}_{\rm min}$ versus ${\cal E}>{\cal E}_{\rm max}$ cases.

\subsection{Geodesic properties}

For a given value of the actions $(\tilde J_r, \tilde J_\theta, \tilde J_\phi)$, one may obtain the constants of the motion $\{{\cal E},{\cal Q},{\cal L}\}$ by inverting the equation for the actions in terms of the constants of the motion.  The determination of ${\cal L}=\tilde J_\phi$ is trivial.  The determination of ${\cal E}$ and ${\cal Q}$ is harder, requiring the solution of a nonlinear system of two equations; we solve these iteratively by first writing a function to obtain ${\cal E}(\tilde J_r,{\cal Q},\tilde J_\phi)$ by bisection solution of $\tilde J_r({\cal E},{\cal Q},{\cal L})=\tilde J_r$; and then writing a function to adjust ${\cal Q}$ (again by a bisection search) until we find the desired $\tilde J_\theta$.

We will often need the $3\times 3$ matrix of partial derivatives
\begin{equation}
{\mathbfss M} =
\left( \begin{array}{ccc}
\frac{\partial {\cal E}}{\partial\tilde J_r} &
\frac{\partial{\cal Q}}{\partial J_r} & \frac{\partial{\cal L}}{\partial\tilde J_r}
\\
\frac{\partial {\cal E}}{\partial\tilde J_\theta} &
\frac{\partial{\cal Q}}{\partial J_\theta} & \frac{\partial{\cal L}}{\partial\tilde J_\theta}
\\
\frac{\partial {\cal E}}{\partial\tilde J_\phi} &
\frac{\partial{\cal Q}}{\partial J_\phi} & \frac{\partial{\cal L}}{\partial\tilde J_\phi}
\end{array} \right).
\end{equation}
The last column of ${\mathbfss M}$ is simply $(0,0,1)^{\rm T}$.  The first column is notable for being the vector of fundamental angular frequencies corresponding to the $r$, $\theta$, and $\phi$ directions on the torus, $(\Omega_r,\Omega_\theta,\Omega_\phi)^{\rm T}$.

It is possible to obtain ${\mathbfss M}$ by numerical differentiation, but it is more accurate to obtain its inverse ${\mathbfss M}^{-1}$ by differentiating the actions with respect to $({\cal E},{\cal Q},{\cal L})$.  The last column (the vector of partial derivatives of $\tilde J_\phi$) is simply $(0,0,1)^{\rm T}$.  The second column (the vector of partial derivatives of $\tilde J_\theta$) can be obtained using the relation
\begin{equation}
\frac{\partial J_\theta}{\partial A} = \frac1{2\pi} \oint \left.\frac{\partial u_z}{\partial A}\right|_z \,\rmd z
= \frac1{4\pi} \oint \left.\frac{\partial( u_z^2 )}{\partial A}\right|_z \,\frac{\rmd z}{u_z},
\label{eq:jtderiv}
\end{equation}
where $A\in\{{\cal E},{\cal Q},{\cal L}\}$, and we have used the fact that $u_z\rightarrow 0$ at the turning points to set to zero terms associated with changes in $z_{\rm min,max}$.  The explicit expressions are
\begin{equation}
u_z = \frac{\bigl[
{\cal Q}(1-z^2) - a^2(1-{\cal E}^2)z^2(1-z^2) - {\cal L}^2z^2
\bigr]^{1/2}}{1-z^2},
\end{equation}
with derivatives
\begin{eqnarray}
\frac{\partial(u_z^2)}{\partial{\cal E}} &=& \frac{2a^2{\cal E} z^2}{1-z^2},
\nonumber \\
\frac{\partial(u_z^2)}{\partial{\cal Q}} &=& \frac1{1-z^2}, {\rm ~~and}
\nonumber \\
\frac{\partial(u_z^2)}{\partial{\cal L}} &=& -\frac{2{\cal L} z^2}{(1-z^2)^2}.
\end{eqnarray}
Near the turning points or for low inclinations, $u_z$ becomes small, which is an issue since it is in the denominator of Eq.~(\ref{eq:jtderiv}).  We thus set $z=z_-\tanh\alpha$, perform the integral for $0<\alpha<\infty$, and then multiply by 4 to get the whole cycle; using Eq.~(\ref{eq:uz2}) this gives
\begin{equation}
\frac{\rmd z}{u_z} = \frac{z_-}{{\cal Q}^{1/2}} \frac{1-z^2}{\sqrt{1-z^2/z_+^2}}\,{\rm sech}\,\alpha\,\rmd\alpha.
\end{equation}
For large inclinations, $z_-/{\cal Q}^{1/2}$ may be obtained directly; for small inclinations ($z_-<10^{-8}$), the equatorial limit may be used,
\begin{equation}
\lim_{{\cal Q}\rightarrow 0^+} \frac{z_-}{{\cal Q}^{1/2}} = \frac1{\sqrt{{\cal L}^2 + a^2(1-{\cal E}^2)}}.
\end{equation}

A similar approach works for the derivatives of the radial action.  In this case, we need
\begin{equation}
\frac{\partial J_r}{\partial A} = \frac1{4\pi} \oint \left.\frac{\partial (u_r^2)}{\partial A}\right|_r \,\frac{\rmd r}{u_r}.
\end{equation}
This time, the desired substitution is Eq.~(\ref{eq:r-beta}), with which we find
\begin{equation}
\frac{\rmd r}{u_r} = \frac{(r_+-r_-)\Delta}{2\sqrt{P(r)}}\,{\rm sech}^2\,\beta\,\rmd\beta,
\end{equation}
where $P(r)$ is the polynomial on the right-hand side of Eq.~(\ref{eq:u-r}).  If we factor the polynomial as
\begin{equation}
P(r) = -(1-{\cal E}^2)(r-r'_-)(r-r'_+)(r-r_-)(r-r_+),
\end{equation}
where $r_\pm$ and $r'_\pm$ are the four roots,\footnote{These are all real in the case of stable orbits since $P(r)$ is negative at $r=0$, positive at the outer horizon $r=r_{{\rm h}+}$, and then has 3 roots outside the outer horizon.} then we may simplify this to
\begin{equation}
\frac{\rmd r}{u_r} = \frac{\Delta}{\sqrt{(1-{\cal E}^2)(r-r'_-)(r-r'_+)}}\,{\rm sech}\,\beta\,\rmd\beta.
\end{equation}
The derivatives are:
\begin{eqnarray}
\frac{\partial(u_r^2)}{\partial {\cal E}} &=& \frac{2[(r^2+a^2){\cal E} - a{\cal L}](r^2+a^2)}{\Delta^2}
+ \frac{2a({\cal L}-a{\cal E})}{\Delta},
\nonumber \\
\frac{\partial(u_r^2)}{\partial {\cal Q}} &=& -\frac1\Delta, {\rm ~~and}
\nonumber \\
\frac{\partial(u_r^2)}{\partial {\cal L}} &=& \frac{-2a[(r^2+a^2){\cal E} - a{\cal L}]}{\Delta^2}
- \frac{2({\cal L}-a{\cal E})}{\Delta}.
\end{eqnarray}

\subsection{Particle position and momentum in terms of the action-angle variables}

In perturbation theory it is critical to be able to obtain the particle's phase space location $(x^i,p_i)$ in terms of the action-angle variables $(\tilde J_i,\psi^i)$.  The generic procedure to do this is as follows.  First, for a given $\{\tilde J_i\}$, we identify the constants of the motion $\{{\cal E},{\cal Q},{\cal L}\}$ on the corresponding torus.  These three actions mutually commute: $\{J_i,J_j\}_{\rm P}=0$, where $\{,\}_{\rm P}$ denotes the Poisson bracket.  Second, we must construct the angle variables.  For actual numerical computation, the method of choice is to use the direct conditions to construct the mapping of $( J_i, \psi^i ) \rightarrow (x^i, p_i)$, which will depend on the (unknown) origin of the angle coordinates $\bpsi=(0,0,0)$ on each torus; and we will find a valid origin by inspection.

We first use the direct conditions \citep[e.g.][Eq.~9.48]{2002clme.book.....G} to write a system of differential equations for $x^i$ and $p_i$ as functions of the angles for fixed ${\bmath J}$:
\begin{equation}
\left.\frac{\partial x^i}{\partial \psi^j}\right|_{\tilde J_k} = \left.\frac{\partial \tilde J_j}{\partial u_i}\right|_{x^k}
{\rm ~and~}
\left.\frac{\partial u_i}{\partial \psi^j}\right|_{\tilde J_k} = -\left.\frac{\partial \tilde J_j}{\partial x^i}\right|_{u_k}.
\label{eq:direct}
\end{equation}
These equations can be re-written in terms of derivatives of constants of the motion,
\begin{equation}
\left.\frac{\partial x^i}{\partial \psi^j}\right|_{\tilde J_k} = \sum_{A\in\{ {\cal E},{\cal Q},{\cal L} \}}
[{\mathbfss M}^{-1}]_{A,j} \left.\frac{\partial A}{\partial u_i}\right|_{x^k},
\label{eq:xpsi}
\end{equation}
and similarly for $u_i$.  These equations define a solution for $\bpsi$, except that we must choose an origin $\bpsi=0$ on each torus; thus all possible solutions differ by a transformation of the form $\psi^i\rightarrow \psi^i + f^i(\tilde{\bmath J})$.

Our next step is to determine an appropriate choice of origin, i.e. the 3-dimensional submanifold of phase space corresponding to $\bpsi={\bmath 0}$.  All valid choices of angle variables correspond to some origin (and are related to each other by simple phase-shifts of the angle variables on each torus), but in multiple dimensions not all origins correspond to valid angle variables.\footnote{A trivial way to see this is to note that the direct conditions show a transformation $\psi^i\rightarrow\psi^i+f^i({\bmath J})$ to be canonical if and only if the $3\times 3$ matrix $\partial f^i/\partial J_j$ is symmetric, i.e. if $f$ is derivable from a potential: $f^i({\bmath J}) = \partial\Phi/\partial J_i$ for some $\Phi({\bmath J})$.}  \citet[][\S50C]{1978mmcm.book.....A} shows that a (locally) valid choice of origin is $Q^i=$constant, where $(Q^i,P_i)$ are a set of canonical coordinates.\footnote{The construction in \citet{1978mmcm.book.....A} technically shows that the generating function for the transformation $(Q^i,P_i)\rightarrow(\psi^i,J_i)$ vanishes at the chosen origin; but inspection shows that $\bpsi={\bmath 0}$ there as well.}  We could thus choose a particular value of $(r,\theta,\phi)$ as our origin; but this would not be applicable to all orbits since there is no value of $r$ that all orbits cross.  We prefer to choose fixed $(p_r,\theta,\phi)$, which is also valid since Hamiltonian mechanics does not distinguish between the position and momentum variables\footnote{This argument is equivalent to applying first a canonical transformation $Q^r=p_r$, $P_r=-r$, and then the construction in \citet{1978mmcm.book.....A}.}; we take $p_r=0$, $\theta=\pi/2$, and $\phi=0$.

It is then necessary only to apply certain inequalities so that each torus intersects the $\bpsi={\bmath 0}$ manifold once and the angle coordinates are defined globally on each torus; we take $\partial H/\partial r<0$ and $p_\theta<0$.  This corresponds to the point of pericentre and ascending node at zero longitude, i.e.
\begin{equation}
x^i = \left(r_-, \frac\pi2, 0\right)
{\rm ~~and~~}
u_i = \left( 0, -{\cal Q}^{1/2} , {\cal L} \right).
\end{equation}

Starting from $\bpsi=(0,0,0)$, we may use Eq.~(\ref{eq:xpsi}) to evolve the particle to any chosen angle coordinates.  Since the construction of the torus integrates over no more than 1 cycle, even a simple integrator is sufficient (we use the 4th order explicit Runge-Kutta method).

We finally need the formulas for the partial derivatives of ${\cal E}$, ${\cal Q}$, and ${\cal L}$ with respect to $(x^i,p_i)$.  For ${\cal E}$, this is simple: the partial derivatives represent the Hamiltonian flow,
\begin{equation}
\frac{\partial {\cal E}}{\partial p_i} = \frac{\partial H}{\partial p_i} = \dot x^i = \frac{u^i}{u^t},
\end{equation}
where $u_t$ is determined from the normalization $g^{\alpha\beta}u_\alpha u_\beta = -1$ and $u^\alpha$ is obtained by raising indices.  The derivatives $\partial{\cal E}/\partial x^i$ can be determined from the conserved quantities, e.g. by taking the $t$-derivative of Eq.~(\ref{eq:u-theta}),
\begin{equation}
2u_\theta \dot u_\theta = [2a^2(1-{\cal E}^2) \cos\theta\sin\theta + 2{\cal L}^2 \cot\theta \csc^2\theta]\dot\theta;
\end{equation}
using that $\dot\theta = u^\theta/u^t = u_\theta/(\Sigma u^t)$, we find
\begin{equation}
-\frac{\partial {\cal E}}{\partial\theta}
= \dot u_\theta = \frac{ a^2(1-{\cal E}^2) \cos\theta\sin\theta + {\cal L}^2 \cot\theta \csc^2\theta }{ \Sigma u^t }.
\end{equation}
We also know trivially that
\begin{equation}
-\frac{\partial {\cal E}}{\partial\phi} = 0.
\end{equation}
Finally, taking $\frac12$ of the $t$-derivative of Eq.~(\ref{eq:u-r}) gives
\begin{eqnarray}
\Delta^2u_r\dot u_r + 2(r-M)\Delta u_r^2 \dot r \!\!\! &=& \!\!\! 2[(r^2+a^2){\cal E}-a{\cal L}] r{\cal E}\dot r
 - r\Delta\dot r
 \nonumber \\ && \!\!\!
 - (r-M)(r^2+K)\dot r.
\end{eqnarray}
One then uses $\dot r = u^r/u^t = \Delta u_r/(\Sigma u^t)$ to obtain:
\begin{eqnarray}
-\frac{\partial{\cal E}}{\partial r} =
\dot u_r \!\!\! &=& \!\!\! \frac{ 2[(r^2+a^2){\cal E}-a{\cal L}] r{\cal E}}{\Delta\Sigma u^t}
- \frac r{\Sigma u^t} 
 \nonumber \\ && \!\!\!
 - \frac{(r-M)(r^2+K)}{\Delta\Sigma u^t} - \frac{2(r-M)}{\Sigma u^t} u_r^2.
\end{eqnarray}

We may find the derivatives of ${\cal Q}$ by taking the differential of Eq.~(\ref{eq:u-theta}):
\begin{eqnarray}
\rmd{\cal Q} \!\!\! &=& \!\!\!
2\bigl[u_\theta\,\rmd u_\theta - a^2(1-{\cal E}^2)\cos\theta\sin\theta\,\rmd\theta
 - a^2{\cal E}\cos^2\theta\,\rmd{\cal E}
\nonumber \\
&& \!\!\!
+ {\cal L}\cot^2\theta\,\rmd{\cal L} - {\cal L}^2 \cot\theta\csc^2\theta\,\rmd\theta\bigr].
\end{eqnarray}
Recalling that ${\cal L}=u_\phi$, and using the aforementioned rules to obtain the partial derivatives of ${\cal E}$, we may find $\partial{\cal Q}/\partial x^i$ and $\partial{\cal Q}/\partial p_i$.

This argument allows us to take any action-angle variables $(\psi^i, J_i)$ and construct the usual coordinates $(x^i,p_i)$.  We have not implemented an inverse function $(x^i,p_i)\rightarrow (\psi^i,J_i)$ since it is not required for this work, although we do not expect it to present any special difficulty.

\section{Gravitational perturbations}
\label{sec:modes}

We next describe the solution of the equations for the Weyl tensor component $\psi_4$ given the particle trajectory.  The approach is to use the separability of the equations to write
\begin{equation}
\psi_4(t,r\theta,\phi) = \rho^4 \sum_{\ell m} \int_{-\infty}^{\infty}
{\cal R}_{\ell m \omega}(r) S^{-2,\chi}_{\ell,m}(\theta) \rme^{\rmi (m\phi-\omega t)} \,\frac{\rmd\omega}
{2\pi},
\label{eq:Psi4}
\end{equation}
where the radial function $R_{\ell m \omega}(r)$ satisfies a homogeneous equation (in vacuum) or an inhomogeneous equation (in the present case, with source).  The separated equation and the behaviour of the radial solutions were considered by \citet{1973ApJ...185..635T}; we will thus describe in detail here only the aspects that are required for either our numerical techniques or for the treatment of the nonrelativistic limit.

\subsection{Angular eigenfunctions}
\label{ss:Stheta}

We are interested here in the solutions of the latitude eigenfunctions $S^{s,\chi}_{\ell,m}(\theta)$ that satisfy the eigenvalue equation
\citep[][Appendix A]{2000PhRvD..61h4004H}
\begin{eqnarray}
-{\cal E}^s_{\ell,m} S^{s,\chi}_{\ell,m}(\theta) \!\!\! &=& \!\!\!
\partial_\theta^2 S^{s,\chi}_{\ell,m}(\theta) + \cot\theta\,\partial_\theta S^{s,\chi}_{\ell,m}(\theta)
\nonumber \\ && \!\!\!
+ \chi^2\cos^2\theta\,S^{s,\chi}_{\ell,m}(\theta) - 2s\chi\cos\theta\,S^{s,\chi}_{\ell,m}(\theta)
\nonumber \\ && \!\!\!
- \frac{m^2+2ms\cos\theta+s^2}{\sin^2\theta}S^{s,\chi}_{\ell,m}(\theta).
\label{eq:S}
\end{eqnarray}
Here $S^{\chi}_{\ell,m}(\theta)$ denotes the values at $\phi=0$; we understand that
\begin{equation}
S^{\chi}_{\ell,m}(\theta,\phi) =
S^{\chi}_{\ell,m}(\theta) \rme^{\rmi m\phi}.
\end{equation}
For gravitational wave problems using the gauge-invariant Weyl tensor component $\psi_4$ one requires the $s=-2$ harmonics with $\chi\equiv a\omega$.  The vertical quantum number $\ell$ begins at $\ell_{\rm min}=\max(|m|,|s|)$ by convention.  The solution method is standard and is described in Appendix~\ref{app:sph}.

\subsection{The radial equation: homogeneous piece}
\label{ss:RrHomo}

The radial equation can be written as, suppressing the indices $\ell m\omega$,
\begin{equation}
\Delta^2 \frac\rmd{\rmd r}\left( \Delta^{-1}\frac{\rmd {\cal R}}{\rmd r}\right) - V{\cal R} = -{\cal T},
\label{eq:R-eq}
\end{equation}
where ${\cal T}$ is a source term to be described later and the potential is \citep[][Eq. 4.9]{1973ApJ...185..635T}
\begin{equation}
V(r) = \frac{-K^2-4\rmi(r-M)K}\Delta + 8\rmi\omega r + {\cal E} - 2am\omega + a^2\omega^2 - 2.
\end{equation}

We consider first the solution of the source-free homogeneous equation subject to either the boundary condition of a purely ingoing gravitational wave at the horizon $r=r_{{\rm h}+}$, or a purely outgoing wave at $r=\infty$.  The matching condition in between in the presence of sources will be considered next.

It is standard to use the radial coordinate $r_\star$ defined by
\begin{equation}
\frac{\rmd r_\star}{\rmd r} = \frac{r^2+a^2}\Delta,
\end{equation}
or explicitly \citep[e.g.][]{2000PhRvD..61h4004H}
\begin{equation}
r_\star = r + \frac{r_{{\rm h}+}}{\sqrt{1-a_\star^2}} \ln \frac{r-r_{{\rm h}+}}{2M} + 
\frac{r_{{\rm h}-}}{\sqrt{1-a_\star^2}} \ln \frac{r-r_{{\rm h}-}}{2M}.
\end{equation}
In the $r_\star$ coordinate, the radial equation becomes
\begin{equation}
\frac{(r^2+a^2)^2}{\Delta} \frac{\rmd^2{\cal R}}{\rmd r_\star^2}
+ 2 \frac{r\Delta - 2(r-M)(r^2+a^2)}{\Delta} \frac{\rmd{\cal R}}{\rmd r_\star} - V{\cal R} = 0.
\end{equation}

\subsubsection{Inner solution}

We consider the inner region first.  In this region, as $r_\star\rightarrow -\infty$ and $r\rightarrow r_{{\rm h}+}$, we have
\begin{equation}
r-r_{{\rm h}+} \approx 2M \exp \left[ \frac{\sqrt{1-a_\star^2}}{r_{{\rm h}+}}(r_\star-r_{{\rm h}+}) \right].
\end{equation}
Then since $\Delta = (r-r_{{\rm h}-})(r-r_{{\rm h}+})$, we have $\Delta\propto \rme^{\Gamma r_\star}$, where
\begin{equation}
\Gamma = \frac{\sqrt{1-a_\star^2}}{r_{{\rm h}+}} = \frac{2\sqrt{M^2-a^2}}{r_{{\rm h}+}^2+a^2}.
\end{equation}
In the last equality we have used the root equation for the horizon, $r_{{\rm h}+}^2-2Mr_{{\rm h}+}+a^2=0$.  We further see that
\begin{equation}
\lim_{r\rightarrow r_{{\rm h}+}} \frac{2(r-M)}{r^2+a^2} = \Gamma
\end{equation}
and
\begin{equation}
\lim_{r\rightarrow r_{{\rm h}+}} \frac K{r^2+a^2} = \omega -m\Omega_{\rm H} \equiv \varpi.
\label{eq:varpidef}
\end{equation}
Then in the limit $r_\star\rightarrow-\infty$, the differential equation becomes
\begin{equation}
\frac{\rmd^2R}{\rmd r_\star^2} - 2\Gamma\frac{\rmd {\cal R}}{\rmd r_\star} + (\varpi^2+2\rmi\Gamma\varpi){\cal R} = 0;
\end{equation}
the two solutions are then exponentials,
\begin{equation}
{\cal R}_1(r_\star) \propto \rme^{(-\rmi\varpi+2\Gamma) r_\star} {\rm ~and~}
{\cal R}_2(r_\star) = \rme^{\rmi\varpi r_\star}.
\end{equation}
\citet{1973ApJ...185..635T} obtained these solutions and found that ${\cal R}_1$ corresponds to the ingoing wave and ${\cal R}_2$ to the outgoing wave.  Thus, interior to any matter sources, the physical solution must be that which matches to $\alpha_1{\cal R}_1$, where $\alpha_1$ is some (possibly complex) constant.  Since $\Gamma>0$ and ${\cal R}_1$ increases exponentially outward relative to ${\cal R}_2$, no numerical difficulty arises in starting at some large negative value of $r_\star$, setting
\begin{equation}
{\cal R}_1 = \Delta^2 \rme^{-\rmi\varpi r_\star} {\rm ~and~}
\frac{\rmd {\cal R}_1}{\rmd r_\star} = (-\rmi\varpi+2\Gamma) \Delta^2 \rme^{-\rmi\varpi r_\star},
\end{equation}
and integrating outward with a standard (RK4) integrator.

\subsubsection{Outer solution}

The radial equation in the outer region ($r_\star\rightarrow\infty$) is not so well behaved.  In this limit, we find
\begin{equation}
\frac{\Delta}{(r^2+a^2)^2}V\rightarrow -\omega^2 + 4\rmi\omega r_\star^{-1} + {\cal O}(r_\star^{-2}).
\end{equation}
and the radial equation becomes (keeping the leading-order terms in $r_\star^{-1}$)
\begin{equation}
\frac{\rmd^2{\cal R}}{\rmd r_\star^2} - \frac2{r_\star} \frac{\rmd {\cal R}}{\rmd r_\star} + \left( \omega^2 - \frac{4\rmi\omega}{r_\star} \right){\cal R} = 0.
\end{equation}
This may be turned into a quadratic equation for the WKB wave number $k$ with the replacement $\rmd/\rmd r_\star\rightarrow \rmi k$; the solutions, to lowest order in $r_\star^{-1}$, are
\begin{equation}
k_3 = \omega - \frac{3\rmi}{r_\star}
{\rm ~~and~~}
k_4 = -\omega + \frac\rmi{r_\star}.
\end{equation}
This implies an imaginary logarithmic divergence of the phases, or equivalently a power-law behavior of the real parts of the solutions at $r_\star\rightarrow\infty$,
\begin{equation}
{\cal R}_3 \rightarrow r_\star^3 \rme^{\rmi\omega r_\star}
{\rm ~~and~~}
{\cal R}_4 \rightarrow r_\star^{-1} \rme^{-\rmi\omega r_\star}.
\label{eq:R3R4}
\end{equation}
Here ${\cal R}_3$ corresponds to a purely outgoing wave and is the physical solution in problems where there is no incident gravitational radiation.  (${\cal R}_4$ corresponds to a purely ingoing wave.)  However, as noted by \citet{1973ApJ...185..649P}, if one integrates from large to small $r_\star$, the ${\cal R}_4$ solution grows relative to ${\cal R}_3$, so it quickly begins to dominate.  Several solutions to this problem exist in the literature, such as using a highly accurate integrator such that the ${\cal R}_4$ solution remains subdominant \citep{1973ApJ...185..649P}; or evolving a linear combination of ${\cal R}$ and $\rmd{\cal R}/\rmd r_\star$ that eliminates the subdominance of ${\cal R}_4$ as $r\rightarrow\infty$ \citep{1973ApJ...185..649P} or lacks the long-range imaginary part of the potential that causes the divergence \citep{1982PhLA...89...68S, 1982PThPh..67.1788S}.

An alternative, which we use here, is to note that Eq.~(\ref{eq:R-eq}) is a regular linear ODE with analytic coefficients except at $r\in\{r_{{\rm h}-},r_{{\rm h}+},\infty\}$.  Therefore, if we desire ${\cal R}_3$ and $\rmd{\cal R}_3/\rmd r$ at any real value of $r>r_{{\rm h}+}$, it is permissible to integrate the ODE on any convenient path through the complex plane.  We note further that if $\Re r$ is large, then while $|{\cal R}_3|$ grows more rapidly than $|{\cal R}_4|$ on the {\em real axis}, if $\Im r$ is allowed to be positive then $|{\cal R}_4|$ is exponentially enhanced relative to $|{\cal R}_3|$.  This suggests that one may integrate not along the real axis itself but along a contour in the first quadrant of the complex plane that begins at large $r$ where an asymptotic solution is valid, and ends on the real axis.\footnote{Since $r\approx r_\star$ in the large-radius regime, we may construct the path of integration in either plane.  Here the $r$-plane is more convenient because we have explicit analytic expressions for the ODE coefficients, so they can be found without writing a routine for the complex function $r(r_\star)$ or expending the substantial computational resources to evaluate such a function at each integration step.}  For concreteness, we note that the ratio of solutions obtained from Eq.~(\ref{eq:R3R4}) should, for large $r_\star$, be
\begin{equation}
\left| \frac{{\cal R}_3}{{\cal R}_4} \right|^{1/2} \propto |r_\star^2 \rme^{\rmi\omega r_\star}|
\propto [(\Re r_\star)^2 + (\Im r_\star)^2] \exp (-\omega\Im r_\star).
\end{equation}
To evaluate ${\cal R}_3$ at some real $r_0$, we integrate along the path
\begin{equation}
\Im r = \min \left\{ \frac3\omega\ln\frac{\Re r}{r_0}, \Re r \right\},
\end{equation}
for which ${\cal R}_4$ dominates as $\Re r\rightarrow\infty$.
In practice, we follow such a path directly to $r_0=r$ if we desire ${\cal R}_3(r)$ at $r>r_{\rm f}\equiv \max\{ |\omega|^{-1}, 3M\}$; for $r<r_{\rm f}$, we integrate first to $r_{\rm f}$ and then leftward along the real axis.  We have experimented with both a complex RK4 integrator and a Bulirsch-Stoer method\footnote{The implementation of the Bulirsch-Stoer method involved taking steps of $\Delta \Re r = -0.03\min(\Re r,\omega^{-1})$.  Each step was computed using the modified midpoint method with $N=4$, 6, 8, and 16 substeps, and extrapolated to $N=\infty$ using a cubic polynomial in $N^{-2}$; see \citet[][\S\S16.3,16.4]{1992nrca.book.....P}.}; we have used the Bulirsch-Stoer integrator here since it is slightly faster for similar accuracy, but we found both methods to be workable.

The starting point for the integration is initialized in accordance with \citet[][Eq. D15]{1973ApJ...185..649P} using terms through order $r^{-2}$ (i.e. $C_2$); our default starting value of $\Re r$ is $1250\max\{M,|\omega|^{-1}\}$.

\subsection{Source term}

We next need the source term ${\cal T}_{\ell m\omega}(r)$ in the Teukolsky equation.  This is given by\footnote{These equations are provided by \citet{1997PThPS.128....1M} and in slightly different form by \citet[][Eq.~4.39]{2000PhRvD..61h4004H}.} as
\begin{equation}
{\cal T}_{\ell m\omega}(r) = \int \rmd t\,{\cal T}_{\ell,m}(r,t) \,\rme^{\rmi\omega t},
\label{eq:Tt}
\end{equation}
where
\begin{eqnarray}
{\cal T}_{\ell,m}(r,t) \!\! &=& \!\!
\Delta^2(r)\big\{ A_0 \delta(r-r_0) + \partial_r[A_1\delta(r-r_0)]
\nonumber \\ && \!\! + \partial_r^2[A_2\delta(r-r_0)] \big\}.
\label{eq:Tr}
\end{eqnarray}
Here $r_0$ denotes the radial coordinate of the particle at time $t$, and the $A$-coefficients are given as follows: for the $\delta$-function,
\begin{eqnarray}
A_0 \!\! &=& \!\!
-2\frac{\Sigma}{\rho^2\Delta^2}[{\opL}_1^\dagger{\opL}_2^\dagger S+ 2\rmi a\rho\sin\theta\,{\opL}_2^\dagger S]C_{nn}
\nonumber \\ && \!\!
+ \frac{2\sqrt2}{\rho^3\Delta}\Bigl[ \left(\frac{\rmi K}\Delta + \frac{2r}{\Sigma}\right){\opL}_2^\dagger S
\nonumber \\ && \!\!
- \left( \frac{\rmi K}\Delta - \frac{2r}{\Sigma}\right)  \frac{2a^2 \sin\theta\cos\theta}{\Sigma} S
\Bigl] C_{n\bar m}
\nonumber \\ && \!\!
+ \frac{\bar\rho}{\rho^3}\left[ \frac{K^2}{\Delta^2} + 2\rmi\rho\frac{K}\Delta + \rmi 
\frac{2\Delta\omega r - 2K(r-M)}{\Delta^2}
\right] S C_{\bar m\bar m};
\nonumber \\ &&
\end{eqnarray}
for the derivative of the $\delta$-function\footnote{There is a spurious factor of $\rho$ in the second term of Eq.~(4.40d) of \citet{2000PhRvD..61h4004H}.},
\begin{eqnarray}
A_1 \!\! &=& \!\!
-\frac{2\sqrt 2}{\rho^3\Delta} \left[ {\opL}_2^\dagger S + \frac{2a^2 \sin\theta\cos\theta}{\Sigma}S \right] C_{n\bar m}
\nonumber \\ && \!\!
+ \frac{2\bar\rho}{\rho^3} \left( \rho - \rmi \frac K\Delta \right)S C_{\bar m\bar m},
\end{eqnarray}
and for the second derivative of the $\delta$-function,
\begin{equation}
A_2 = -\frac{\bar\rho}{\rho^3}SC_{\bar m\bar m},
\end{equation}
where we have suppressed the arguments of the spheroidal harmonic $S\equiv S_{\ell,m}^{-2,a\omega}(\theta,\phi)$.  The coefficients of the stress-energy tensor are
\begin{equation}
C_{ab} = \mu \frac{({\bmath u}\cdot{\bmath a})({\bmath u}\cdot{\bmath b})}{\Sigma u^t},
\end{equation}
where ${\bmath a}$ and ${\bmath b}$ are null vectors (either ${\bmath n}$ or $\bar{\bmath m}$).
The values of $\opL_1^\dagger\opL_2^\dagger S$ and $\opL_2^\dagger S$ can be obtained from Eqs.~(\ref{eq:L2d}) and (\ref{eq:L1L2d}).

Now for a quasiperiodic trajectory along the torus, we may write ${\cal T}_{\ell,m}(r,t)$ as a function of the angle variables,
${\cal T}_{\ell,m}(r,t) = {\cal T}_{\ell,m}[r,\bpsi(t)]$,
where each $\psi_i(t)$ advances at the rate $\dot\psi_i=\Omega_i$.  Then we take the Fourier transform,
\begin{equation}
{\cal T}_{\ell,m}(r,\bpsi) = \sum_{\bmath q} {\cal T}_{\ell,m}(r|{\bmath q}) \rme^{-\rmi{\bmath q}\cdot{\bpsi}},
\end{equation}
where ${\bmath q}$ is a lattice vector (i.e. $q_r$, $q_\theta$, and $q_\phi$ are all integers).  Using $\bpsi = \bpsi^{(0)} + {\bmath\Omega}t$, we may integrate Eq.~(\ref{eq:Tt}) to get:
\begin{equation}
{\cal T}_{\ell m\omega}(r) = \sum_{\bmath q} {\cal T}_{\ell,m}(r|{\bmath q}) \rme^{\rmi{\bmath q}\cdot\bpsi^{(0)}} \,2\pi\delta(\omega-{\bmath\Omega}\cdot{\bmath q}).
\label{eq:Tt2}
\end{equation}
With Eq.~(\ref{eq:Tt2}), we may evolve each value of ${\bmath q}$ separately, treating ${\cal T}_{\ell,m}(r|{\bmath q})$ as the source, and then sum the resulting perturbations.

The Fourier components ${\cal T}_{\ell,m}(r|{\bmath q})$ may be evaluated as follows.  We first see that
\begin{equation}
{\cal T}_{\ell,m}(r|{\bmath q}) = \int \frac{\rmd^3\bpsi}{(2\pi)^3} \,{\cal T}_{\ell,m}(r,\bpsi) \rme^{\rmi{\bmath q}\cdot{\bpsi}}.
\label{eq:tcomplex}
\end{equation}
Now if we increment $\psi^\phi$ by some amount $\delta\psi^\phi$, then it is easy to see that $\phi$ is increased by $\delta\psi^\phi$ while the other phase space coordinates $\{r,\theta,u_r,u_\theta,u_\phi\}$ remain fixed.  Thus ${\cal T}_{\ell,m}(r,\bpsi)$ is multiplied by $\exp(-\rmi m\delta\psi^\phi)$.  Since the complex exponential in Eq.~(\ref{eq:tcomplex}) is multiplied by $\exp(\rmi q_\phi\delta\psi_\phi)$, it follows that ${\cal T}_{\ell,m}(r|{\bmath q})$ is nonzero only if $q_\phi=m$.  In this case, the $\psi^\phi$ integral is also trivial, so we find
\begin{equation}
{\cal T}_{\ell,m}(r|{\bmath q}) = \delta_{m,q_\phi} \int \frac{\rmd\psi^r\,\rmd\psi^\theta}{(2\pi)^2} \,{\cal T}_{\ell,m}(r,\bpsi) \rme^{\rmi{\bmath q}\cdot{\bpsi}}.
\label{eq:tcomplex2}
\end{equation}
This provides a means of computing ${\cal T}_{\ell,m}(r|{\bmath q})$ while doing only a double integral over the torus instead of a triple integral.  In practical computation, the integral is computed as a discretized sum over $N_rN_\theta$ equally spaced points on the $\psi^\phi=0$ subtorus.
This completes the approximation of ${\cal T}_{\ell,m}(r|{\bmath q})$ by a finite sum over $\delta$-functions and their derivatives.

\subsection{Solution to the inhomogeneous radial Teukolsky equation}
\label{ss:inhomo}

We solve the full radial Teukolsky equation via a Green's function method.  The starting point is to recognize that given the boundary conditions, the solution must satisfy
\begin{equation}
{\cal R}(r) = \left\{\begin{array}{lll}
Z^{\rm down} {\cal R}_1(r) & & r<r_1 \\
Z^{\rm out} {\cal R}_3(r) & & r>r_1
\end{array}\right.,
\end{equation}
where $Z^{\rm down,~out}$ are undetermined constants.  We now suppose that the source contained a $\delta$-function at some radius $r_1$, i.e. we had an inhomogeneous equation,
\begin{equation}
\Delta^2\frac\rmd{\rmd r}\left( \Delta^{-1}\frac{\rmd{\cal R}}{\rmd r}\right) - V{\cal R} = -\delta(r-r_1).
\end{equation}
This would imply the jump conditions that $R$ be continuous at $r_1$ and that its derivative jump by
\begin{equation}
{\cal R}'(r_1+\epsilon) - {\cal R}'(r_1-\epsilon) = \frac1{\Delta(r_1)}.
\end{equation}
These two conditions allow us to solve for $Z^{\rm down,~out}$:
\begin{equation}
Z^{\rm down} = \frac{{\cal R}_3(r_1)}{\Delta(r_1)W_{31}(r_1)} {\rm ~~and~~}
Z^{\rm out} = \frac{{\cal R}_1(r_1)}{\Delta(r_1)W_{31}(r_1)},
\end{equation}
where the Wronskian is
\begin{equation}
W_{31}(r) = {\cal R}_3(r){\cal R}'_1(r) - {\cal R}_1(r) {\cal R}'_3(r).
\end{equation}
The Wronskian of the two solutions to a second-order ODE may be obtained by elementary means: in this case, we have $W(r)\propto\Delta(r)$, so we write $W_{31}(r)=\aleph \Delta(r)$.  An evaluation at one point is sufficient to determine $\aleph$.

We thus have the full solution in the interior region ($r<r_-$)
\begin{equation}
{\cal R}_{\ell m\omega}(r) = \sum_{\bmath q} Z^{\rm down}_{\ell m,{\bmath q}} {\cal R}_1(r) \rme^{\rmi{\bmath q}\cdot\bpsi^{(0)}}\,2\pi\delta(\omega-{\bmath q}\cdot{\bmath\Omega}),
\end{equation}
where integration of the Green's function gives
\begin{equation}
Z^{\rm down}_{\ell m,{\bmath q}} = \aleph_{\ell m\omega}^{-1} \int [ A_0{\cal R}_3 - A_1{\cal R}'_3 + A_2{\cal R}''_3 ] \, \rme^{\rmi{\bmath q}\cdot\bpsi} \,
\frac{\rmd^3\bpsi}{(2\pi)^3},
\label{eq:Green}
\end{equation}
and $A_0$ and ${\cal R}_3$ are evaluated at the particle position.  (The $A_1$ and $A_2$ terms are obtained similarly using integration by parts to move the radial derivative from the argument of ${\cal T}$ to the argument of the Green's function.)  A similar equation is valid in the exterior region $r>r_+$ for the outgoing wave amplitude $Z^{\rm out}_{\ell m,{\bmath q}}$ if we swap ${\cal R}_1\leftrightarrow {\cal R}_3$.

Using Eq.~(\ref{eq:Psi4}), it follows that in the interior region,
\begin{equation}
\psi_4 = \rho^4 \sum_{\ell m, {\bmath q}} Z^{\rm down}_{\ell m,{\bmath q}} {\cal R}_1(r) S_{\ell,m}^{-2,a\omega}(\theta)\rme^{\rmi m\phi}
\rme^{\rmi{\bmath q}\cdot\bpsi^{(0)}}
\rme^{-\rmi\omega t},
\label{eq:interior}
\end{equation}
and in the exterior region
\begin{equation}
\psi_4 = \rho^4 \sum_{\ell m, {\bmath q}} Z^{\rm out}_{\ell m,{\bmath q}} {\cal R}_3(r) S_{\ell,m}^{-2,a\omega}(\theta)\rme^{\rmi m\phi}
\rme^{\rmi{\bmath q}\cdot\bpsi^{(0)}}
\rme^{-\rmi\omega t}.
\label{eq:exterior}
\end{equation}
At large radii, $\rho^4{\cal R}_3\rightarrow r^{-1}\rme^{\rmi\omega r_\star}$.  Then, since the flux of gravitational waves at large radii is the time-average of $|\psi_4|^2/(4\pi\omega^2)$, we may integrate over the sphere (using $\int |S|^2\,\sin\theta\,\rmd\theta\,\rmd\phi=2\pi$) to get the emitted power to $\infty$:
\begin{equation}
P_{\rm out} = \sum_{\ell,m,{\bmath q}} \frac{|Z^{\rm out}_{\ell m,{\bmath q}}|^2}{2\omega^2},
\end{equation}
where $\omega = {\bmath q}\cdot{\bmath\Omega}$.
The power emitted into the black hole was derived by \citet{1974ApJ...193..443T}; the solution is
\begin{equation}
P_{\rm down} = \sum_{\ell,m,{\bmath q}} \alpha \frac{|Z^{\rm down}_{\ell m,{\bmath q}}|^2}{2\omega^2},
\end{equation}
where
\begin{equation}
\alpha = \frac{8192M^5r^5_{{\rm h}+}\varpi(\varpi^2+\Gamma^2)(\varpi^2+4\Gamma^2)\omega^3}{|C|^2}.
\end{equation}
Here $C$ is the Starobinsky-Teukolsky coefficient, whose squared absolute value is
\begin{eqnarray}
|C|^2 \!\!\! &=&  [(\lambda+2)^2+4m\chi -4\chi^2](\lambda^2+36m\chi-36\chi^2)
\nonumber \\ && \;
+ 48(2\lambda+3)\chi(2\chi-m)+144\omega^2(M^2-a^2),
\end{eqnarray}
and we have used $\chi=a\omega$ and $\varpi=\omega-m\Omega_{\rm H}$.\footnote{Note that \citet[][Eq. 4.18]{2000PhRvD..61h4004H} contains a missing factor of $m$ in the first term; $[(\lambda+2)^2+4a\omega_{mk}-4a^2\omega^2_{mk}]$ should read $[(\lambda+2)^2+4ma\omega_{mk}-4a^2\omega^2_{mk}]$.  Also note that ``$\epsilon$'' as defined in \citet{1974ApJ...193..443T} is $\Gamma/2$ here.}

The energy and angular momentum radiated (both to infinity and into the hole) are required in order to follow the evolution of circular {\em or} equatorial orbits under radiation reaction \citep[e.g.][]{1978ApJ...225..687D, 1993PhRvD..48..663S, 1994PhRvD..50.6297S, 1998PhRvD..58f4012K, 2000PhRvD..61h4004H}; comparison of $\dot{\cal E}$ and $\dot{\cal L}$ to literature values can be used a test of our code.  Evolution of generic orbits that are both eccentric and inclined would also require a relation for $\dot{\cal Q}$ \citep{2003PhRvD..67h4027M, 2005PhRvL..94v1101H, 2006PhRvD..73b4027D}, which is not required for this paper.

We have tested our code by checking our computed energy and angular momentum fluxes against the results from Table~VI of \citet{2006PhRvD..73b4027D}, for $M=1$, $a_\star=0.9$, semilatus rectum $p\equiv 2/(r_-^{-1}+r_+^{-1})=6$, and a range of eccentricities $e$ [defined by $r_+/r_-=(1+e)/(1-e)$] and inclinations $\theta_{\rm inc} \equiv \pi/2-\theta_{\rm min}$.  We consider all modes with $\max\{\ell,|q_\phi|,|q_\theta|,|q_r|\}\le j_{\rm max}$, and expect convergence as $j_{\rm max}\rightarrow\infty$.  Comparisons are given in Table~\ref{tab:compare}.

\begin{table*}
\caption{\label{tab:compare}Comparison of our energy and angular momentum fluxes at $\infty$ and at the horizon to those of \citet{2006PhRvD..73b4027D} (DH) for $M=1$, $a_\star=0.9$, and semilatus rectum $6$.  The ``Error'' column gives the maximum fractional error  of any of the four columns relative to DH.}
\begin{tabular}{cccccccc}
\hline\hline
$e$ & $\theta_{\rm inc}$ & $j_{\rm max}$ & $\dot{E}^{\rm H}/\mu^2$ & $\dot{E}^\infty/\mu^2$ & $\dot{L}^{\rm H}/\mu^2$ & $\dot{L}^\infty/\mu^2$ & Error
\\
\hline
0.1 & 20$^\circ$ & 4 & $-$4.2574E$-$6 & $+$5.8126E$-$4 & $-$6.7238E$-$5 & $+$8.4497E$-$3 & 1.1E$-$2 \\
& & 6 & $-$4.2576E$-$6 & $+$5.8700E$-$4 & $-$6.7241E$-$5 & $+$8.5310E$-$3 & 6.8E$-$4 \\
& & 8 & $-$4.2576E$-$6 & $+$5.8738E$-$4 & $-$6.7241E$-$5 & $+$8.5362E$-$3 & 3.6E$-$5 \\
& & DH & $-$4.2576E$-$6 & $+$5.8740E$-$4 & $-$6.7241E$-$5 & $+$8.5365E$-$3 & \\
\hline
0.3 & 40$^\circ$ & 4 & $-$5.8169E$-$6 & $+$7.0118E$-$4 & $-$1.0006E$-$4 & $+$7.6189E$-$3 & 3.7E$-$2 \\
& & 6 & $-$5.8857E$-$6 & $+$7.2361E$-$4 & $-$1.0061E$-$4 & $+$7.8091E$-$3 & 4.4E$-$3 \\
& & 8 & $-$5.8882E$-$6 & $+$7.2636E$-$4 & $-$1.0063E$-$4 & $+$7.8316E$-$3 & 5.8E$-$4 \\
& & DH & $-$5.8882E$-$6 & $+$7.2678E$-$4 & $-$1.0063E$-$4 & $+$7.8350E$-$3 & \\
\hline\hline
\end{tabular}
\end{table*}

\section{The resonant amplitude}
\label{sec:ResonantAmp}

Having now solved for $\psi_4$, it remains to compute the resonant amplitude ${\cal S}^{(m)}$ from Paper I.  While it would in principle be possible to compute the metric perturbation directly, by constructing the master potential $\Psi$ \citep{1978PhRvL..41..203W, 2003PhRvD..67l4010O} and then utilizing the \citet{1975PhRvD..11.2042C} procedure, we will find it more useful to express ${\cal S}^{(m)}$ directly in terms of $\psi_4$.

Furthermore, since we are considering Lindblad resonances, the metric perturbations are required only in the interior and exterior regions, i.e. at radii $r<r_-$ or $r>r_+$, where the vacuum Einstein equation is obeyed.  This will simplify our task greatly.

The key to the computation of the resonant amplitude is the result from Paper I that
\begin{equation}
{\cal S}^{(m)} = \frac{2\rmi}{m\Omega_{\rm s}\mu_1{\cal Z}\epsilon} {\cal P}^{(m)},
\label{eq:SP}
\end{equation}
where ${\cal P}^{(m)}$ is the power provided by the $m$ Fourier mode of the metric perturbation to a test particle of mass $\mu_1\rightarrow 0$ on an orbit that is slightly eccentric, oscillating between $R-\epsilon$ and $R+\epsilon$, where $\epsilon$ is small.

The power can be computed without direct knowledge of the metric perturbations, but it breaks into two similar cases for the ILRs and OLRs.  In both cases, we use the fact that knowledge of $\psi_4$ in a neighborhood around the test particle's radius enables determination of the metric perturbations (up to gauge modes and to the zero-frequency ``$\ell=0$ and 1 modes'' corresponding to changes in the mass and spin of the hole, which provide no power) and hence the power is the same as that which would be provided by a pure gravitational wave solution with the same $\psi_4$.

The perturber in our case is on a circular equatorial orbit, hence $\tilde J_r=\tilde J_\theta=0$ and no $q_r,q_\theta\neq 0$ need be considered.  The mode of interest has $q_\phi=m$, $\omega=m\Omega_{\rm s}$, and pattern speed $\Omega_{\rm s} = \Omega_\phi$ (evaluated at the perturber position).  Without loss of generality, we set the initial longitude $\psi^{(0)}_\phi=0$.

\subsection{Inner Lindblad resonances}

In the case of an ILR, the Weyl tensor component $\psi_4$ is given by Eq.~(\ref{eq:interior}).  This is exactly the same as the case of an incoming gravitational wave with azimuthal quantum number $m$ and frequency $\omega = m\Omega_{\rm s}$ with the specified amplitudes in each $\ell$ mode.  In such a situation, one may see that the radial mode is
\begin{equation}
{\cal R}_{\ell m\omega}(r) = Z^{\rm down}_{\ell m} {\cal R}_1(r) = Z^{\rm down}_{\ell m} [c_{13}{\cal R}_3(r) + c_{14}{\cal R}_4(r) ],
\end{equation}
where $c_{13}$ and $c_{14}$ are constants evaluated in Appendix~\ref{app:wronskian}.  The power in incoming gravitational waves, outgoing waves, and waves going down into the hole are given by
\begin{eqnarray}
P_{\rm in} \!\! &=& \!\! \sum_{\ell m} \frac{(2\omega)^8}{|C|^2}\,\frac{|c_{14}Z^{\rm down}_{\ell m}|^2}{2\omega^2},
\nonumber \\
P_{\rm out} \!\! &=& \!\! \sum_{\ell m} \frac{|c_{13}Z^{\rm down}_{\ell m}|^2}{2\omega^2}, {\rm ~and}
\nonumber \\
P_{\rm down} \!\! &=& \!\! \sum_{\ell m} \frac{\alpha|Z^{\rm down}_{\ell m}|^2}{2\omega^2}.
\end{eqnarray}
Now we consider our test particle.  It too emits gravitational waves, including a set of modes at azimuthal quantum number $m$ and at the frequency
\begin{equation}
\omega = m\Omega(R) - \kappa = m\Omega_{\rm s}.
\end{equation}
These waves are emitted both down into the hole and out to infinity, with amplitudes $Z^{\rm down}_{1,\ell m}$ and $Z^{\rm out}_{1,\ell m}$ that are calculable by the same procedure as for the perturber, but this time with Fourier modes $(q_r,q_\theta,q_\phi)=(-1,0,m)$.

We may now obtain the power absorbed by the test particle using conservation of energy.  There is a correction to the power escaping to $\infty$ and down the black hole in accordance with
\begin{eqnarray}
\delta P_{\rm out} \!\!\! &=& \!\!\! \Re \sum_{\ell m} \frac{c_{13}Z^{\rm down}_{\ell m} Z^{{\rm out}\ast}_{1,\ell m}}{\omega^2}
{\rm ~~and~~}
\nonumber \\
\delta P_{\rm down} \!\!\! &=& \!\!\! \Re \sum_{\ell m} \frac{\alpha Z^{\rm down}_{\ell m} Z^{{\rm down}\ast}_{1,\ell m}}{\omega^2}.
\label{eq:P1}
\end{eqnarray}
The power absorbed by the test particle is the negative of this, which can be found by expanding the real part as one-half the sum of a quantity and its complex conjugate:
\begin{eqnarray}
\sum_{m\in\mathbb Z} {\cal P}^{(m)} \!\!\! &=& \!\!\! -\delta P_{\rm out} - \delta P_{\rm down}
\nonumber \\
&=& \!\!\! -\frac1{2\omega^2} \sum_{\ell m} Z^{\rm down}_{\ell m} \left(
c_{13} Z^{{\rm out}\ast}_{1,\ell m} + \alpha Z^{{\rm down}\ast}_{1,\ell m}
\right)
\nonumber \\ &&
\!\!\! -\frac1{2\omega^2} \sum_{\ell m} Z^{{\rm down}\ast}_{\ell m} \left(
c_{13}^\ast Z^{{\rm out}}_{1,\ell m} + \alpha Z^{{\rm down}}_{1,\ell m}
\right).
\end{eqnarray}
We may identify the individual contributions ${\cal P}^{(m)}$ by noting that it is linear in the $m$ Fourier mode of the metric perturbation; and thus it arises from the terms proportional to $Z^{\rm down}_{\ell m}$ or $Z^{{\rm down}\ast}_{\ell,- m}$.\footnote{Since $\psi_4$ is a complex quantity whose real and imaginary parts encode different components of the Weyl tensor, perturbations in the metric tensor, curvature, etc. are not linear in $\psi_4$ alone but rather are linear in $\psi_4$ and $\psi_4^\ast$.  Thus the $m$ Fourier mode of the metric perturbation depends on both the $m$ and $-m$ Fourier modes of $\psi_4$.}  Therefore:
\begin{eqnarray}
{\cal P}^{(m)} \!\!\! &=& \!\!\! -\frac1{2\omega^2} \sum_\ell Z^{\rm down}_{\ell m} \left(
c_{13} Z^{{\rm out}\ast}_{1,\ell m} + \alpha Z^{{\rm down}\ast}_{1,\ell m}
\right)
\nonumber \\ &&
\!\!\! -\frac1{2\omega^2} \sum_\ell Z^{{\rm down}\ast}_{\ell, -m}\! \left(
c_{13-}^\ast Z^{{\rm out}}_{1,\ell,- m}\! + \alpha_- Z^{{\rm down}}_{1,\ell,- m}\!
\right)\!.
\label{eq:P2}
\end{eqnarray}
Here $c_{13-}$ refers to the coefficient for negative values of $m$ and $\omega$: $c_{13-}(\ell,m,\omega)\equiv c_{13}(\ell,-m,-\omega)$, and similarly for $\alpha_-$ (note that the $\alpha$-coefficients are real).  Inspection of the radial equation shows that $c_{13-}^\ast=c_{13}$ and $\alpha_-=\alpha$.  In the particular case where both the perturber and the test particle are in the equatorial plane, there also exists a reflection symmetry of the emitted waveform across the equator, e.g. $Z^{{\rm down}\ast}_{\ell, -m} = (-1)^m Z^{\rm down}_{\ell m}$.  Therefore the two terms in Eq.~(\ref{eq:P2}) are equal.  Thus we see that the power absorbed by the test particle in all of the frequency $\omega$ modes is
\begin{equation}
{\cal P}^{(m)} = -\frac1{\omega^2}\sum_\ell Z^{\rm down}_{\ell m} \left(
c_{13} Z^{{\rm out}\ast}_{1,\ell m} + \alpha Z^{{\rm down}\ast}_{1,\ell m}
\right).
\label{eq:P-ILR}
\end{equation}
This has the correct dependences: it is manifestly linear in $\mu_1$, which is essential since the computation of the resonant amplitude requires division by $\mu_1$, and also it is linear in the epicyclic oscillation amplitude $\epsilon$ since the order $q_r$ Fourier mode of the gravitational wave scales as $\epsilon^{|q_r|}$.

\subsection{Outer Lindblad resonances}

A related argument applies to the OLRs.  This time, we consider a perturber on a circular orbit, again emitting at frequency $\omega = m\Omega_{\rm s}$, and a test particle on a slightly eccentric orbit emitting at frequency
\begin{equation}
\omega = m\Omega(R) + \kappa,
\end{equation}
i.e. we are considering the $(q_r,q_\theta,q_\phi)=(1,0,m)$ Fourier mode on its torus.  This time, since we are considering a vacuum solution outside the perturber's orbit, the perturber (or at least its $m\neq0$ part) may be replaced by a gravitational wave coming out of the hole's past horizon.  The radial mode amplitude is now
\begin{equation}
{\cal R}_{\ell m\omega}(r) = Z^{\rm out}_{\ell m} {\cal R}_3(r) = Z^{\rm down}_{\ell m} [c_{31}{\cal R}_1(r) + c_{32}{\cal R}_2(r) ].
\end{equation}

The changes in power escaping to infinity and going down into the hole are now
\begin{eqnarray}
\delta P_{\rm out} \!\!\! &=& \!\!\! \Re\sum_{\ell m} \frac{Z^{\rm out}_{\ell m} Z^{{\rm out}\ast}_{1,\ell m}}{\omega^2}
{\rm ~~and~~}
\nonumber \\
\delta P_{\rm down} \!\!\! &=& \!\!\! \Re\sum_{\ell m} \frac{\alpha c_{31} Z^{\rm out}_{\ell m} Z^{{\rm down}\ast}_{1,\ell m}}{\omega^2};
\end{eqnarray}
but we note that Eq.~(\ref{eq:c31}) implies $\alpha c_{31}=-c_{13}^\ast$.
The power absorbed by the test particle from the $m$ Fourier mode of the metric perturbation is now
\begin{equation}
{\cal P}^{(m)} = -\frac1{\omega^2}\sum_\ell Z^{\rm out}_{\ell m} \left(
 Z^{{\rm out}\ast}_{1,\ell m} - c_{13}^\ast Z^{{\rm down}\ast}_{1,\ell m}
\right).
\label{eq:P-OLR}
\end{equation}

Equations~(\ref{eq:P-ILR}) and (\ref{eq:P-OLR}) at first appear remarkable: they show that the torques at the Lindblad resonances, which depend on ${\cal S}^{(m)}$, can be related to the overlap between the gravitational waveforms emitted by the perturber and a test particle at the location of the resonance.  But this could have been expected: the same time-dependent multipole moments that are responsible for the gravitational wave emission also generate resonant torques.

We are now ready to compute the resonant amplitudes ${\cal S}^{(m)}$.  We consider three cases.  First we review the case of a Keplerian disc, showing how the Lindblad torques can be treated via the Teukolsky formalism.  Then we consider a disc around a Schwarzschild black hole with a perturber, similar to the physical situation envisaged by \citet{2009arXiv0906.0825C}; this is the first case for which the relativistic machinery developed in Paper I and here is actually necessary, and we find an additional $m=1$ ILR with no Newtonian Keplerian analogue.\footnote{The new ILR does however exist for any Newtonian potential with an ISCO.}  Finally, we compute the resonance strengths in the case of an equatorial orbit around a Kerr black hole.

\section{Resonances in the nonrelativistic limit}
\label{sec:Kepler}

The problem of Lindblad resonance torques in Newtonian Keplerian discs (i.e. discs in nonrelativistic motion around a central point mass with negligible pressure gradient) has been treated many times; here we treat it using the Teukolsky equations.  We wish to find $|{\cal S}^{(m)}|^2$ for each resonance.  This requires us first to find $Z^{\rm out,down}_{\ell m,{\bmath q}}$ for both circular orbits (the perturber) and slightly eccentric orbits (for the test particle).  We work at radii $\gg M$.  The solutions for the radial Teukolsky functions in this regime are described in Appendix~\ref{app:R0}; the angular functions are simply the spin-weighted spherical harmonics.  As is well-known, the Lindblad resonances can be found at values of the test particle radius
\begin{equation}
r_1 = \left( \frac {m\mp1}m \right)^{2/3}r_0 \equiv \varsigma^{\mp}_m r_0,
\end{equation}
where the upper and lower signs refer to the inner and outer Lindblad resonances.

\subsection{Emitted waves: circular orbit}

We consider first a particle on a circular Keplerian orbit at radius $r_0\gg M$, orbiting at angular velocity $\Omega_\phi=M^{1/2}r_0^{-3/2}$.  The required stress-energy coefficients phased to zero longitude are
\begin{equation}
C_{nn}=\frac{\mu}{4r_0^2}, {\rm~~}
C_{n\bar m} = \frac{-\rmi \mu M^{1/2}}{2\sqrt2\,r_0^{5/2}}, {\rm ~~and~~}
C_{\bar m\bar m} = -\frac{\mu M}{2r_0^3}.
\end{equation}
The leading-order source term is then
\begin{equation}
A_0 = -\frac\mu{2r_0^2}\opL_1^\dagger\opL_2^\dagger S\left(\theta=\frac\pi2\right).
\end{equation}
(the $A_1$ and $A_2$ terms have powers of $r_0^{-3/2}$ and $r_0^{-1}$ respectively; when they are integrated, the additional $\partial_r$ or $\partial_r^2$ makes these subdominant to $A_0$).  We will find it convenient to define
\begin{eqnarray}
y_{\ell m} \!\!\! &\equiv& \!\!\! \opL_1^\dagger\opL_2^\dagger S\left(\theta=\frac\pi2\right)
\nonumber \\ &=&\!\!\! \sqrt{2\pi\frac{(\ell+2)!}{(\ell-2)!}}
 \; Y_{\ell m}\left(\theta=\frac\pi2,\phi=0\right)
\end{eqnarray}
so that $A_0=-\mu y_{\ell m}/(2r_0^2)$.  

Now for the circular orbit, a particular $m$-mode is excited only at $\omega = m\Omega_\phi = mM^{1/2}r_0^{-3/2}$, and the Fourier mode of the torus that excites it is $(q_r,q_\theta,q_\phi)=(0,0,m)$.  The downward and outward radiation amplitudes are obtained from Eq.~(\ref{eq:Green}), with the formulae for ${\cal R}_1$, ${\cal R}_3$, and $\aleph$ from Appendix~\ref{app:R0}:
\begin{eqnarray}
Z^{\rm down}_{\ell m;0,0,m} \!\!\!&=&\!\!\! \frac{{\cal R}_3(r_0)A_0}\aleph = \frac{\mu y_{\ell m}}{2(2\ell+1)k_1r_0^{\ell+1}}
{\rm~~and}\nonumber\\
Z^{\rm out}_{\ell m;0,0,m} \!\!\!&=&\!\! \!\frac{{\cal R}_1(r_0)A_0}\aleph = \rmi^{2-\ell}\frac{(\ell-2)!\,\mu y_{\ell m}(2\omega)^{\ell+2}r_0^\ell}{2\cdot(2\ell+1)!}.
\label{eq:BKc}
\end{eqnarray}
Note the $\propto r_0^{-(\ell+1)}$ and $\propto r_0^\ell$ radial behaviour; this is expected for sourcing the order-$\ell$ multipole.

\subsection{Emitted waves: eccentric orbit}

We now consider a test particle of mass $\mu_1$ orbiting at radius $r_1$, and with slight eccentricity $\epsilon/r_1$ such that the particle oscillates between $r_1-\epsilon$ and $r_1+\epsilon$.  We are now interested in the $(\mp1,0,m)$ Fourier mode (where as in Paper I the upper sign represents the ILR and the lower sign the OLR), which has frequency $\omega = (m\mp1)M^{1/2}r_1^{-3/2}$.  As this is a resonance we will not distinguish between this value of $\omega$ and that for the perturber.

The computation of $A_0$ and negligibility of $A_{1,2}$ proceed in an exactly analogous way to that for the circular orbit; the only differences are that (i) the true radius $r$ differs from its mean value $r_1$; and (ii) we must now work at general longitude since we no longer have trivial angle integrals. We find
\begin{equation}
A_0 = -\frac{\mu_1 y_{\ell m}}{2r^2}\rme^{-\rmi m\phi}.
\end{equation}
The amplitude emitted to future null infinity is
\begin{equation}
Z^{\rm down}_{1;\ell m;\mp1,0,m} = \int_0^{2\pi} \frac{R_3(r)A_0}\aleph \rme^{\mp\rmi \psi^r} \frac{\rmd\psi_r}{2\pi},
\end{equation}
where the integrand may be evaluated at $\psi^\phi=0$ since the $\psi^\phi$ integral is trivial. The waveform emitted into the future horizon $Z^{\rm out}_{\ell m;\mp 1,0,m}$ may be obtained by replacing ${\cal R}_3(r)$ with ${\cal R}_1(r)$.

The epicyclic motion in the Kepler potential can be found in any dynamics text \citep[e.g.][]{2000ssd..book.....M}; expressed in our variables, it is, at $\psi^\phi=0$,
\begin{equation}
r = r_1 - \epsilon\cos\psi^r, {\rm~~and~~}
\phi = 2\frac\epsilon{r_1}\sin\psi^r.
\end{equation}
To first order in $\epsilon$, we then have
\begin{equation}
\int_0^{2\pi} r^n \rme^{-\rmi m\phi} \rme^{\mp\rmi \psi^r} \frac{\rmd\psi^r}{2\pi}
= \left( -\frac n2 \mp m\right) \epsilon r_1^{n-1}.
\end{equation}
Therefore, we conclude that
\begin{eqnarray}
Z^{\rm down}_{1;\ell m;\mp1,0,m} \!\!\!&=&\!\!\! \frac{\mu_1\epsilon y_{\ell m}}{2(2\ell+1)k_1r_1^{\ell+2}} 
\left( \frac{\ell+1}{2}\mp m\right)
{\rm ~~and}\nonumber \\
Z^{\rm out}_{1;\ell m;\mp1,0,m} \!\!\!&=&\!\!\! \rmi^{2-\ell}\frac{(\ell-2)!\,\mu_1\epsilon y_{\ell m}(2\omega)^{\ell+2}r_1^{\ell-1}}{2\cdot(2\ell+1)!}
\nonumber \\ && \times
\left(-\frac{\ell}{2}\mp m\right).
\label{eq:BKe}
\end{eqnarray}

\subsection{Resonant amplitudes}

We are now ready to evaluate Eqs.~(\ref{eq:P-ILR}) and (\ref{eq:P-OLR}), each of which has two terms.  We focus on the ILRs; the treatment of the OLRs is analogous.  A comparison of the two terms shows that, using Eq.~(\ref{eq:BKe}) and the relations in Appendix~\ref{app:R0},
\begin{equation}
\left|\frac{\alpha Z^{\rm down}_{1;\ell m;\mp1,0,m}}{c_{13} Z^{\rm out}_{1;\ell m;\mp1,0,m}}\right| \sim |\varpi|M \left(\frac{M}{r_1}\right)^{2\ell+1} \ll 1,
\end{equation}
so the $Z^{\rm out}_{1;\ell m;\mp1,0,m}$ term dominates in Eq.~(\ref{eq:P-ILR}).  The actual evaluation using Eq.~(\ref{eq:BKc}) as well gives
\begin{equation}
{\cal P}^{(m)} = -\rmi\mu \mu_1 \epsilon\omega \sum_{\ell=m}^\infty
\frac{(\ell-2)!}{(\ell+2)!}
 \frac{(\ell+2m)y_{\ell m}^2}{2\ell+1} \frac{r_1^{\ell-1}}{r_0^{\ell+1}}.
\label{eq:P}
\end{equation}

The summation in Eq.~(\ref{eq:P}) can be simplified using:
\begin{eqnarray}
\frac{(\ell-2)! y_{\ell m}^2}{(\ell+2)! (2\ell+1)}  \!\!\! &=& \!\!\!
\frac1{4\pi} \int_0^{2\pi} P_\ell(\cos\phi) \cos(m\phi) \,\rmd\phi
\nonumber \\
&=& \!\!\! \frac1{4\pi(\ell!)}\left.\frac{\rmd^\ell}{\rmd\varsigma^\ell}
\int_0^{2\pi} \frac{ \cos(m\phi) \,\rmd\phi }{ \sqrt{ 1 + \varsigma^2 - 2\varsigma\cos\phi } }
\right|_{\varsigma=0}
\nonumber \\
&=& \!\!\! 
\frac1{4(\ell!)}\left.\frac{\rmd^\ell}{\rmd\varsigma^\ell} b_{1/2}^{(m)}(\varsigma) \right|_{\varsigma=0}.
\end{eqnarray}
Here the first equality arises by considering the spherical harmonic addition theorem, applying it to points on the equator at longitudes $0$ and $\phi$, and taking the Fourier transform over $\phi$; the second from the generating function relation for the Legendre polynomials; and the third from the definition of the Laplace coefficient.  With this, and using the Taylor expansion formula (and the fact that the Taylor series of $b_{1/2}^{(m)}$ begins with the order $\varsigma^m$ term for $m\ge 0$), we find
\begin{equation}
{\cal P}^{(m)} = -\rmi\frac{\mu \mu_1 \epsilon\omega}{4r_0^2} \left[ b_{1/2}^{(m)}{'}(\varsigma_m^-) + \frac{2 mb_{1/2}^{(m)}(\varsigma_m^-)}{\varsigma_m^- }\right],
\end{equation}
where here the $'$ on the Laplace coefficient denotes differentiation with respect to the argument.  It follows that
\begin{equation}
{\cal S}^{(m)} = \frac{\mu \omega}{2m\Omega_{\rm s}{\cal Z} \varsigma_m^- r_0^2} [ \varsigma_m^- b_{1/2}^{(m)}{'}(\varsigma_m^-) + 2 mb_{1/2}^{(m)}(\upsilon_m^-)].
\end{equation}
The prefactor simplifies using $\Omega_{\rm s}=M^{1/2}/r_0^{3/2}$ and $\omega=m\Omega_{\rm s}$, leaving us with
\begin{equation}
{\cal S}^{(m)} = \frac{qM^{1/2}(\varsigma_m^-)^{1/2}}{2r_0^{1/2}} [ \varsigma_m^- b_{1/2}^{(m)}{'}(\varsigma_m^-) + 2 mb_{1/2}^{(m)}(\varsigma_m^-)].
\end{equation}
This is equivalent to the result from Paper I using the Newtonian potential $h_{tt}$.

For the OLRs, a similar argument holds: the $c_{13}^\ast Z^{{\rm down}\ast}_{1,\ell m}$ term dominates over $Z^{\rm out}_{1,\ell m}$ in Eq.~(\ref{eq:P-OLR}), yielding
\begin{equation}
{\cal P}^{(m)} = -\rmi\mu\mu_1\epsilon\omega \sum_{\ell=\max\{m,2\}}^\infty
\frac{(\ell-2)!}{(\ell+2)!}
\frac{(\ell+1+2m)y_{\ell m}^2}{2\ell+1} \frac{ r_0^\ell }{ r_1^{\ell+2} }.
\end{equation}
We then repeat the conversion of the summation to a Taylor series, this time using the identity $b_{1/2}^{(m)}(\varsigma^{-1})=\varsigma b_{1/2}^{(m)}(\varsigma)$ to relate the series in powers of $r_0/r_1$ to the Laplace coefficient at $r_1/r_0$.  This gives
\begin{eqnarray}
{\cal S}^{(m)} \!\!\! &=& \!\!\! \frac{qM^{1/2}(\varsigma_m^+)^{1/2}}{2r_0^{1/2}} \Bigl[ -\varsigma_m^+ b_{1/2}^{(m)}{'}(\varsigma_m^-) + 2 mb_{1/2}^{(m)}(\varsigma_m^+)
\nonumber \\ &&
- 4(\varsigma_m^+)^{-2} b_{1/2}^{(1)}{'}(0)\delta_{m1}\Bigr],
\label{eq:smolr}
\end{eqnarray}
where the last term arises for $m=1$ because the summation over modes begins at $\ell=2$, whereas the Taylor series of $b_{1/2}^{(1)}$ has a first-order term, $b_{1/2}^{(1)}{'}(0) = 1$.  This can be compared to the result for Paper I, where the last term was $-\varsigma_m^+\delta_{m1}$.  The two terms are exactly equal at resonance $\varsigma_1^+=2^{2/3}$; recall that the resonance is however the only location where ${\cal S}^{(m)}$ is needed.  Indeed, if one does a Newtonian calculation of ${\cal S}^{(m)}$ but working in the inertial frame (where the indirect term in the disturbing function is replaced by a term corresponding to the displacement of the primary), then one derives the last term in Eq.~(\ref{eq:smolr}) in the form presented here.  Of course, the two forms are equivalent on resonance as guaranteed by the gauge invariance arguments of Paper I.

\section{Resonances in the Schwarzschild problem}
\label{sec:Schwarzschild}

We now come to our the first case where we explicitly compute angular momentum transport coefficients in a black hole spacetime: the Schwarzschild system.  We first present the background coefficients and resonance locations, and then give the amplitudes.  To simplify our expressions and avoid proliferation of ``$r/M$'', we will use units where the mass of the black hole is $M=1$.

\subsection{Circular orbits: a review}

For circular orbits at radius $r$, the specific angular momentum and energy of a circular orbit are \citep[][\S19$b$i$\alpha$]{1992mtbh.book.....C}
\begin{equation}
{\cal L} = \frac r{\sqrt{r-3}} {\rm ~~and~~}
{\cal E} = \frac{r-2}{\sqrt{r(r-3)}}.
\end{equation}
Their derivatives are
\begin{equation}
{\cal L}' = \frac{r-6}{2(r-3)^{3/2}} {\rm ~~and~~}
{\cal E}' = \frac{r-6}{2r^{3/2}(r-3)^{3/2}}.
\end{equation}
The angular velocity is
\begin{equation}
\Omega = \frac{{\cal E}'}{{\cal L}'} = r^{-3/2}.
\end{equation}
The conversion from proper to coordinate time is
\begin{equation}
w^t = \frac{\rmd t}{\rmd \tau} = \sqrt{\frac r{r-3}}.
\end{equation}
The epicyclic frequency is
\begin{equation}
\kappa = \frac{\sqrt{r-6}}{r^2},
\end{equation}
and the specific epicyclic impedance is
\begin{equation}
{\cal Z} = \frac1{r-2}\sqrt{\frac{r-6}{r(r-3)}}.
\end{equation}
We see that the epicyclic frequency and impedance both vanish at the ISCO $r=r_{\rm ISCO}=6$.

We now suppose that a perturber is placed on a circular equatorial orbit at radius $r_{\rm s}>r_{\rm ISCO}=6$.  Lindblad resonances of azimuthal quantum number $m$ occur at
\begin{equation}
D(r) = m(r^{-3/2} - r_{\rm s}^{-3/2}) \mp \frac{\sqrt{r-6}}{r^2} = 0.
\label{eq:D-sch}
\end{equation}
There is no simple closed-form solution to this equation.  However, we can deduce its properties by noting that
\begin{equation}
D'(r) = \frac32r^{-5/2}\left[ -m \pm \frac{r-8}{\sqrt{r(r-6)}} \right].
\label{eq:D-deriv-sch}
\end{equation}
Since $(r-8)/\sqrt{r(r-6)}<1$, it follows that $D'(r)<0$ for all positive $m$ and $r>r_{\rm ISCO}$.  Thus we see that for each type of resonance (ILR or OLR) and for a given value of $m$, there is at most one solution to Eq.~(\ref{eq:D-sch}).  Furthermore, we easily see that $D>0$ for $r\approx r_{\rm ISCO}$ and $D<0$ at $r=\infty$, so there exists exactly one ILR and one OLR for each positive integer $m$.

Here we note a key difference from the Newtonian Keplerian case: there exists an $m=1$ ILR.  Ordinarily, the innermost Lindblad resonance is the $m=2$ ILR (mean motion ratio 2:1), in which the test particle goes through two epicyclic periods in every synodic period.  Due to pericentre precession, the Schwarzschild metric admits the $m=1$ ILR, in which the orbital frequency of the perturber is equal to the pericentre precession frequency of the test particle.  This is {\em not} a uniquely relativistic phenomenon, but can occur in any system whose attractive potential at small $r$ exhibits a steeper than $r^{-1}$ dependence, e.g. the potential in the equatorial plane of an oblate planet.  Indeed, there is a ringlet of Saturn at 1.29 Saturn radii, whose pericentre precession rate nearly matches the orbital frequency of Titan, and which has thus acquired a large forced eccentricity \citep{1984Icar...60....1P}.

\subsection{Resonance strengths}

We may now compute ${\cal S}^{(m)}$ by the method of Sec.~\ref{sec:ResonantAmp} for each of the resonances.  The first three ILRs are displayed in Figure~\ref{fig:ilrplot}, where we plot the resonance location $r_1$ as a function of the secondary location $r_0$; and also the torque strength with the perturbing mass and disk density normalized out,
\begin{equation}
N = \frac T{2\pi r_1 q^2\Sigma(r_1)} = \mp\frac{\pi m w^t{\cal Z}}{|D'(r_1)|}|{\cal S}^{(m)}|^2.
\label{eq:Tnorm}
\end{equation}
The normalized resonance strength as measured by $N$ has the advantage of converging to a constant in the Newtonian Keplerian limit, i.e. as $r_0\rightarrow\infty$, for the resonances that exist in this case ($m\ge 2$ ILRs and all OLRs). Its departure from constant behaviour is indicative of relativistic effects.

The resonance positions and strengths are tabulated in Table~\ref{tab:ilrS}.
The maximum value of $\ell$ used in the computation is a balance between computation time and overflow avoidance versus accuracy.  At very large $\ell$ and small $\omega$, the determination of e.g. $\aleph$ and $c_{13}$ are susceptible to overflow errors due to the power-law behaviour with large indices ($r^{1-\ell}$ and $r^{2+\ell}$) of the radial solutions to the Teukolsky equation between $r\sim 2$ and $r\sim \omega^{-1}$.\footnote{In principle such errors could be removed by working with $\ln{\cal R}(r)$ instead of ${\cal R}(r)$, but we have not done this as it would have resulted in much more complex code (including branching to avoid numerical instabilities when ${\cal R}$ passes near zero).  An alternative would have been to define a new floating-type data type with more bits in the exponent.}  Fortunately, for the results in this paper we do not need to work in a regime where overflow occurs.  For most cases, have used $\ell_{\rm max}=20$ for the compuations at $20<r_0\le 250$ and $\ell_{\rm max}=40$ at $8\le r_0\le 20$.\footnote{The exceptions are that for $20<r_0\le40$ we use $\ell_{\rm max}=40$ for the $m=3$ ILRs; and for $r_0>20$ we use $\ell_{\rm max}=30$ for the $m=2$ OLRs.}  For the $m=1$, 2, and 3 ILRs presented, we have estimated the truncation error in $\ell$ by extrapolating\footnote{Since there is a strong odd-even pattern to the contributions from successive multipoles, we used the last two even $\ell$s to generate a geometric sequence of even $\ell$s and did a similar independent procedure for the odd $\ell$s.} the sequence of contributions from successive $\ell$; such errors are found to be $\le 0.1$\% ($m=1$ and $m=2$) and $\le 1$\% ($m=3$).


\subsubsection{The $m\ge 2$ ILRs}

\begin{figure*}
\includegraphics[angle=-90,width=6.5in]{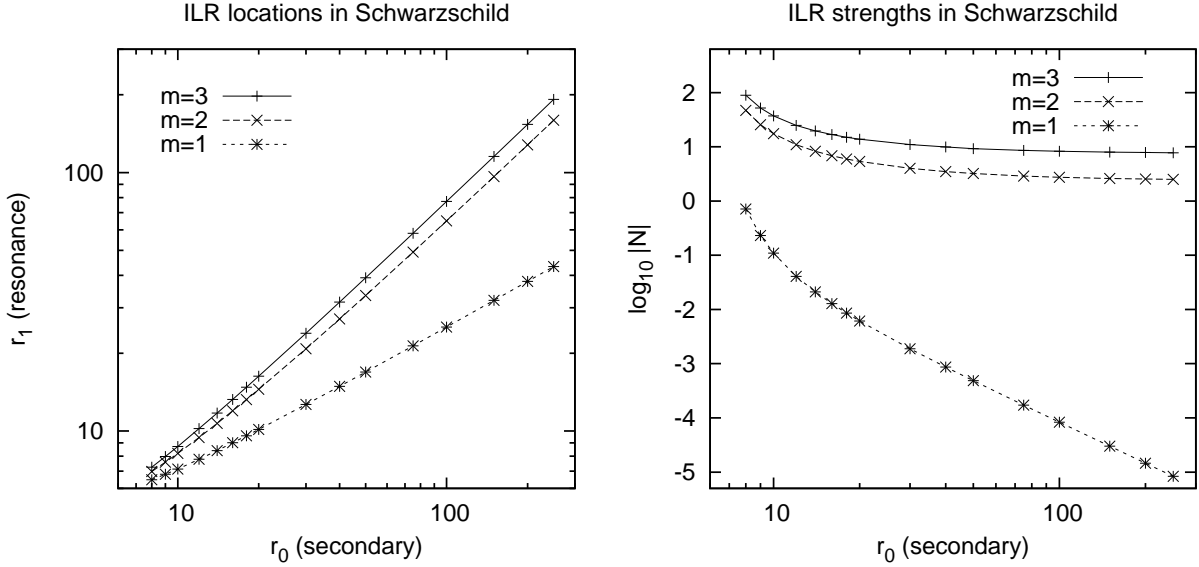}
\caption{\label{fig:ilrplot}The locations and strengths of the inner Lindblad resonances in the Schwarzschild spacetime, shown on logarithmic axes.  The $m\ge 2$ resonances have Newtonian Keplerian analogues; their strength in these units approaches a constant in the Newtonian Keplerian regime, but grows rapidly as one approaches the ISCO.  The $m=1$ resonance exists only due to the pericentre precession, and is found at much smaller radius; it is also much weaker, although its strength grows as we move inward.}
\end{figure*}

The $m\ge 2$ ILRs exist in the Newtonian Keplerian limit as $(m-1):m$ mean motion resonances, and are located at a fixed ratio of semimajor axes, or in this case, orbital radii:
\begin{equation}
\lim_{r_0\rightarrow\infty} \frac{r_1}{r_0} = \left( \frac{m-1}m \right)^{2/3} = \left\{\begin{array}{lll}
0.63 & & m=2 \\ 0.76 & & m=3.
\end{array}
\right.
\end{equation}
These formulae would correspond in the left panel of Fig.~\ref{fig:ilrplot} to straight lines with unit slope (since this is a log-log plot). In fact they are relatively good approximations even at modest values of $r_0$: for the $m=2$ ILR, for example, $r_1/r_0$ increases from 0.63 ($r_0=\infty$) to 0.67 ($r_0=50$) to 0.72 ($r_0=20$). As the secondary approaches the ISCO, however, the resonance locations must remain between the secondary and the ISCO, and hence
\begin{equation}
\lim_{r_0\rightarrow r_{\rm ISCO}} \frac{r_1}{r_0} = 1.
\end{equation}
This behaviour can be seen in the left panel of Fig.~\ref{fig:ilrplot}, where all of the resonance location curves converge to the point $(r_0,r_1)=(6,6)$. Of course, for any finite mass ratio, the assumptions used throughout this paper of weak perturbations and a thin disc would break down before this point is reached.

The resonant strength (as measured by $N$) approaches a constant in this limit,
\begin{equation}
\lim_{r_0\rightarrow\infty} N = - \frac{\pi}{6} (\varsigma_m^-)^2 \left| \varsigma_m^- b_{1/2}^{(m)}{'}(\varsigma_m^-) + 2 mb_{1/2}^{(m)}(\varsigma_m^-) \right|^2.
\end{equation}
This evaluates to $-2.36$ for $m=2$ and $-7.50$ for $m=3$; the convergence to these constant values can be seen from the right panel of Fig.~\ref{fig:ilrplot}.  As one moves inward toward the ISCO, the strength $|N|$ increases. The qualitative effect is unsurprising since the resonance locations become closer to the secondary. It is however noteworthy that the $m\ge 2$ ILR strengths are enhanced substantially relative to the Newtonian Keplerian limit even at large distances from the black hole: the deviation is already 10 per cent at $r_0=160$, and reaches a factor of 2 at $r_0=25$.

\begin{table*}
\caption{\label{tab:ilrS}The resonant strengths in the Schwarzschild problem for $m=1$ and $m=2$ Lindblad resonances; here $r_0$ is the orbital radius of the perturber and $r_1$ is the resonance location.  Truncation errors due to choice of $\ell_{\rm max}$ are estimated to be $\le0.1$\% for the cases given in the table.}
\begin{tabular}{rrrrrrrrrrrrrrrrr}
\hline\hline
 & & \multicolumn{3}{c}{\mbox{$m=1$ ILR (0:1)}} & & \multicolumn{3}{c}{\mbox{$m=2$ ILR (1:2)}}
 & & \multicolumn{3}{c}{\mbox{$m=2$ OLR (3:2)}} & & \multicolumn{3}{c}{\mbox{$m=1$ OLR (2:1)}} \\
$r_0$ & & $r_1$ & & $N$ & & $r_1$ & & $N$ & & $r_1$ & & $N$ & & $r_1$ & & $N$ \\
\hline
8.00 & & 6.48 & & $-$7.11E$-$1 & & 6.97 & & $-$4.69E$+$1 & & 9.55 & & $+$2.28E$+$1 & & 11.33 & & $+$7.05E$-$2 \\
9.00 & & 6.80 & & $-$2.32E$-$1 & & 7.57 & & $-$2.55E$+$1 & & 10.91 & & $+$1.79E$+$1 & & 12.99 & & $+$8.76E$-$2 \\
10.00 & & 7.13 & & $-$1.10E$-$1 & & 8.19 & & $-$1.74E$+$1 & & 12.26 & & $+$1.51E$+$1 & & 14.62 & & $+$1.01E$-$1 \\
12.00 & & 7.78 & & $-$4.08E$-$2 & & 9.44 & & $-$1.09E$+$1 & & 14.92 & & $+$1.30E$+$1 & & 17.85 & & $+$1.19E$-$1 \\
14.00 & & 8.40 & & $-$2.10E$-$2 & & 10.70 & & $-$8.27E$+$0 & & 17.57 & & $+$1.18E$+$1 & & 21.07 & & $+$1.31E$-$1 \\
16.00 & & 9.01 & & $-$1.28E$-$2 & & 11.97 & & $-$6.85E$+$0 & & 20.21 & & $+$1.11E$+$1 & & 24.26 & & $+$1.40E$-$1 \\
18.00 & & 9.58 & & $-$8.57E$-$3 & & 13.23 & & $-$5.96E$+$0 & & 22.84 & & $+$1.07E$+$1 & & 27.46 & & $+$1.46E$-$1 \\
20.00 & & 10.14 & & $-$6.12E$-$3 & & 14.50 & & $-$5.37E$+$0 & & 25.47 & & $+$1.04E$+$1 & & 30.65 & & $+$1.51E$-$1 \\
30.00 & & 12.67 & & $-$1.89E$-$3 & & 20.82 & & $-$4.00E$+$0 & & 38.60 & & $+$9.60E$+$0 & & 46.56 & & $+$1.66E$-$1 \\
40.00 & & 14.89 & & $-$8.69E$-$4 & & 27.13 & & $-$3.49E$+$0 & & 51.72 & & $+$9.30E$+$0 & & 62.45 & & $+$1.73E$-$1 \\
50.00 & & 16.91 & & $-$4.86E$-$4 & & 33.44 & & $-$3.22E$+$0 & & 64.83 & & $+$9.13E$+$0 & & 78.33 & & $+$1.77E$-$1 \\
75.00 & & 21.36 & & $-$1.72E$-$4 & & 49.21 & & $-$2.89E$+$0 & & 97.59 & & $+$8.94E$+$0 & & 118.03 & & $+$1.82E$-$1 \\
100.00 & & 25.25 & & $-$8.33E$-$5 & & 64.97 & & $-$2.75E$+$0 & & 130.36 & & $+$8.85E$+$0 & & 157.72 & & $+$1.85E$-$1 \\
150.00 & & 32.00 & & $-$3.01E$-$5 & & 96.47 & & $-$2.61E$+$0 & & 195.88 & & $+$8.77E$+$0 & & 237.10 & & $+$1.87E$-$1 \\
200.00 & & 37.91 & & $-$1.46E$-$5 & & 127.98 & & $-$2.54E$+$0 & & 261.40 & & $+$8.73E$+$0 & & 316.47 & & $+$1.88E$-$1 \\
250.00 & & 43.25 & & $-$8.35E$-$6 & & 159.48 & & $-$2.50E$+$0 & & 326.92 & & $+$8.70E$+$0 & & 395.84 & & $+$1.89E$-$1 \\
\hline\hline
\end{tabular}
\end{table*}

\subsubsection{The $m=1$ ILR}

For the $m=1$ resonance, the strength is however much less, especially in the nearly Newtonian regime.  This is in part due to the location of the resonance, with $r_1\ll r_0$, and also due to the fact that the Newtonian quadrupole tidal field does not contribute to ${\cal S}^{(m)}$: reflection symmetry across the equatorial plane allows only $m\in\{-2,0,2\}$ contributions to the tidal field, and so the lowest-order contribution to the resonance strength comes from the (gravitoelectric) octupole ($\ell=3$).

While the $m=1$ ILR does not exist in the Newtonian Keplerian problem, its location and strength may be estimated in the large-$r_0$ limit. The $m=1$ ILR location is determined by the condition that the pericentre precession rate,
\begin{equation}
\Omega-\kappa = \frac{\sqrt{r_1}-\sqrt{r_1-6}}{r_1^2} \approx \frac{3}{r_1^{5/2}},
\end{equation}
correspond to the secondary orbital angular velocity, $r_0^{-3/2}$.  This implies, for large $r_0$,
\begin{equation}
r_1 \approx 3^{2/5}r_0^{3/5} \approx 1.55r_0^{3/5}.
\label{eq:r1m1}
\end{equation}
One can see this behaviour in the left panel of Fig.~\ref{fig:ilrplot}: Eq.~(\ref{eq:r1m1}) predicts that the $m=1$ ILR location curve should be a straight line with slope $\frac35$, which is indeed correct at large $r_0$. The deviation from this expression is only 8 per cent at $r_0=20$, which is remarkable.

The strength of the resonance in the large-$r_0$ limit can be estimated from Eq.~(\ref{eq:P}); the leading-order term is $\ell=3$, which gives
\begin{equation}
{\cal S}^{(1)} \approx 4.36r_0^{-19/10} {\rm~~and~~}
N \approx -19.2r_0^{-13/5}.
\end{equation}
This result is valid at very large $r_0$.  However, at even modest $r_0$ it substantially overestimates the strength of the $m=1$ resonance: the true $N$ is smaller by a factor of 0.75 at $r_0=250$ and 0.66 at $r_0=50$.  The principal reason is that there is another contribution to ${\cal P}^{(1)}$ from the gravitomagnetic quadrupole mode ($\ell=2$, negative parity), which does not exist in the Newtonian theory but has the correct symmetry properties for two equatorial orbits to interact via an $m=1$ mode.  Roughly speaking, the gravitomagnetic interaction should give a contribution to ${\cal S}^{(1)}$ that is suppressed by the product of the orbital velocities $v_0v_1 \sim r_0^{-1/2}r_1^{-1/2}\sim r_0^{-4/5}$, but (due to the angular momentum barrier for $\ell=2$ versus 3) enhanced relative to the gravitoelectric octupole by a factor of $(r_1/r_0)^{-1}\sim r_0^{2/5}$.  Thus overall, the gravitomagnetic quadrupole interaction is only weaker than the gravitoelectric octupole by a factor of $\sim r_0^{-2/5}$.  It turns out that the two contributions to ${\cal S}^{(1)}$ have opposite sign, resulting in a suppression of the $m=1$ ILR strength.  The correction is not small:
\begin{equation}
\frac{{\cal S}^{(1)}({\rm magnetic~quadrupole})}{{\cal S}^{(1)}({\rm electric~octupole})} =
\left\{\begin{array}{lll} -0.19, & \!\!r_0=250 \\ -0.38, & \!\!r_0=50, \end{array}\right.
\end{equation}
and then the resonant torque depends on the square of ${\cal S}^{(m)}$ so these corrections are effectively doubled.

The reason for the opposite sign of the gravitomagnetic quadrupole contribution can be understood from linearized gravity arguments.  To lowest order, a moving particle in the vicinity of a moving perturber experiences a gravitomagnetic ``acceleration'' \citep[][\S4.4a]{1984ucp..book.....W}:
\begin{equation}
{\bmath a} = -4{\bmath v}\times{\bmath B}, ~~~~
{\bmath B}({\bmath r}) = qM{\bmath v}_0\times \frac{{\bmath r} - {\bmath r}_0}{|{\bmath r} - {\bmath r}_0|^3},
\end{equation}
i.e. ${\bmath B}$ is the field generated from the momentum in the same way that a magnetic field is generated by electric current.  Here ${\bmath r}_0$ is the position of the perturber and ${\bmath v}_0$ is its velocity.  The test particle experiences an inward gravitomagnetic acceleration that is strongest at inferior conjunction (i.e. when the longitudes of the test particle and perturber are equal).  This is the opposite of the Newtonian gravitoelectric octupole field, which produces an outward force at inferior conjunction.

\subsubsection{The OLRs}

The outer Lindblad resonances, being external to the perturber, are more similar to their Newtonian counterparts than the inner Lindblad resonances.  The limiting strengths as $r_0\rightarrow\infty$ for the $m=1$ (2:1) and $m=2$ (3:2) OLRs are $N=0.19$ and $N=8.62$ respectively; their behaviour at smaller radii is shown in Table~\ref{tab:ilrS}.

For the strong $m=2$ OLR, the resonant strength increases as we move inward because the Lindblad resonances are closer to the perturber than they are in the Newtonian Keplerian case.  However, the weaker $m=1$ OLR (2:1) suffers from the same partial cancellation of gravitoelectric octupole and gravitomagnetic quadrupole contributions as the $m=1$ ILR.  Therefore at small radii it actually becomes weaker.

\section{Resonances in the Kerr problem}
\label{sec:Kerr}

We may now move on to the resonances associated with the circular, equatorial orbits in the Kerr spacetime.  Again, we use units where the mass of the primary hole is $M=1$, and hence $a=a_\star$.  We consider orbits with $\dot\phi>0$; thus $a>0$ (prograde spin) refers to the case where the disc orbit and black hole spin are in the same direction, and $a<0$ (retrograde spin) refers to the opposite case.  The machinery we have developed in the previous sections is completely general and may be used to compute resonance strengths in Kerr with no new difficulties.

The problem is very similar to that of the Schwarzschild spacetime: there exists an ISCO at which $\kappa\rightarrow 0$, and hence once again there exists an $m=1$ ILR.  This time the basic frequencies are
\begin{equation}
\Omega = \frac1{r^{3/2} + a} {\rm ~~and~~}
\kappa = \Omega\sqrt{1 - \frac6r + \frac{8a}{r^{3/2}} - \frac{3a^2}{r^2}}
\end{equation}
\citep[][Appendix]{1987PASJ...39..457O}.  The sign of the $a$ term in $\kappa/\Omega$ implies that pericentre precession is enhanced for $a<0$; the same effect is responsible for the larger value of $r_{\rm ISCO}$ for retrograde spin.

\begin{figure*}
\includegraphics[angle=-90,width=6.5in]{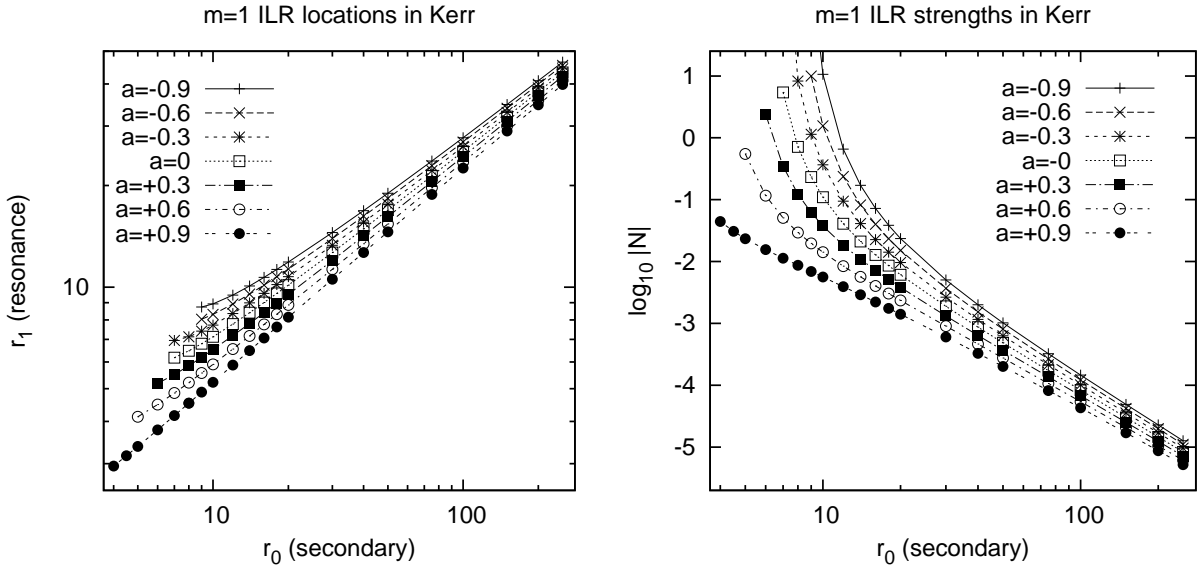}
\caption{\label{fig:kerrm1}The locations (left) and strengths (right) of the $m=1$ ILR for equatorial orbits in the Kerr spacetime. The points show computations using our perturbation theory code, with the symbols indicating the choice of primary spin $a$.  For prograde orbits ($a>0$) the resonance moves inward and become weaker, whereas for retrograde orbits ($a<0$) the resonance moves outward and becomes stronger.}
\end{figure*}

In Fig.~\ref{fig:kerrm1}, we explore the location and strength of $m=1$ ILR as a function of the secondary (perturber) location $r_0$ and the spin of the primary $a$. We would intuitively expect that retrograde spin ($a<0$) would both move the resonance location $r_1$ outward and increase its strength. This expectation is confirmed numerically. Moreover, the effect is quite strong: even at $r_0=250$, a spin of $|a|=0.9$ leads to a factor of 1.17 difference in the $m=1$ ILR location depending on the direction of the spin ($r_1=39.8$ for prograde, 46.4 for retrograde) and a factor of 2.4 in the strength $|N|$ ($5.2\times 10^{-6}$ for prograde, $1.3\times 10^{-5}$ for retrograde). The difference in resonant strength between prograde and retrograde configurations becomes greater as $r_0$ moves inward, and at $r_0=20$ and $|a|=0.9$ is more than an order of magnitude.

At very small radii, we once again have the behaviour $r_1\rightarrow r_0$ and $|N|\rightarrow\infty$ as $r_0\rightarrow r_{\rm ISCO}$. This behaviour is present but not obvious in Fig.~\ref{fig:kerrm1} because $r_{\rm ISCO}$ depends on $a$ (it is larger for the retrograde configuration).

\begin{figure*}
\includegraphics[angle=-90,width=6.5in]{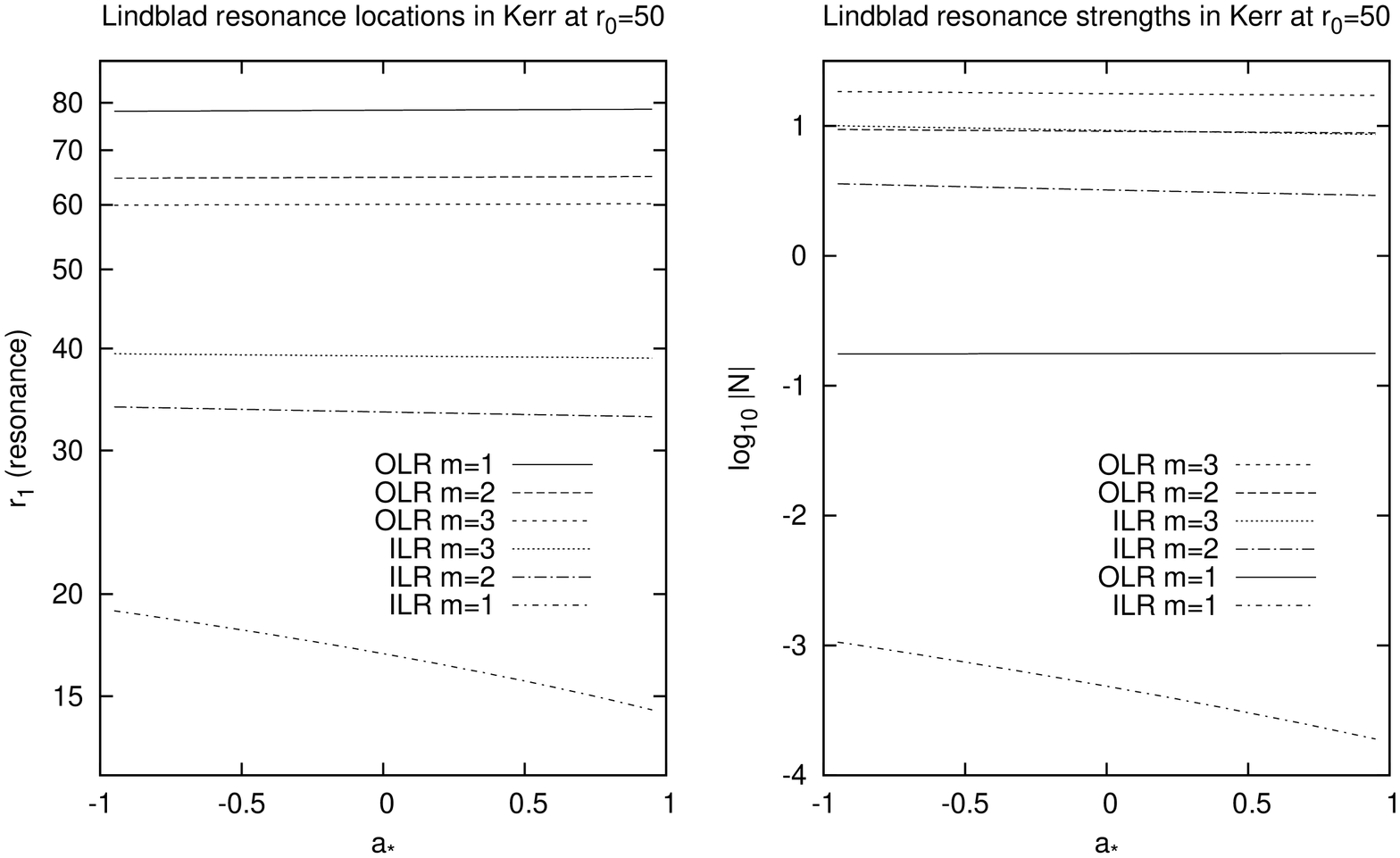}
\caption{\label{fig:lkerr}The locations and strengths of the Lindblad resonances as a function of black hole spin for a perturber in a circular orbit at $r_0=50M$.}
\end{figure*}

\begin{figure*}
\includegraphics[angle=-90,width=6.5in]{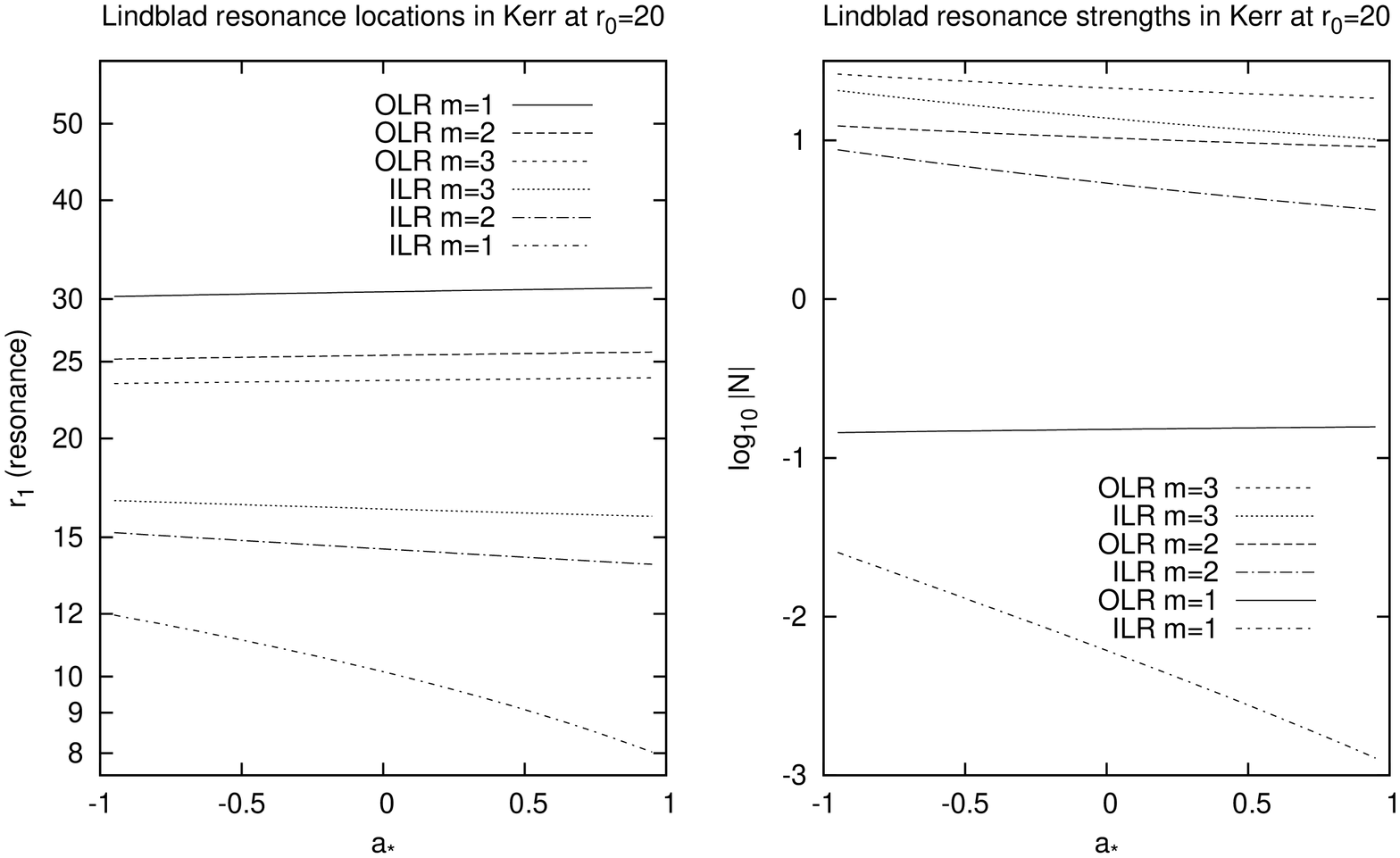}
\caption{\label{fig:lkerr20}The locations and strengths of the Lindblad resonances as a function of black hole spin for a perturber in a circular orbit at $r_0=20M$.}
\end{figure*}

The variation of the Lindblad resonance locations and strengths at fixed $r_0$ but varying $a$ is displayed in Fig.~\ref{fig:lkerr} for $r_0=50$ and Fig.~\ref{fig:lkerr20} for $r_0=20$. For the retrograde spins all of the resonances move closer to the perturber, and correspondingly they are strengthened.  However, we can see that the effect is strongest for the $m=1$ ILR, which is unsurprising since it is closest to the hole and therefore most affected by spin.



\section{Discussion}
\label{sec:discussion}

The Newtonian formulae for the torque applied to a disc at the Lindblad resonances associated with a perturber on a circular equatorial orbit have been extended into the relativistic regime.  The calculation has revealed both new physical effects, and has provided a mathematical connection between seemingly disparate phenomena: resonant torques and gravitational radiation.

At the {\em physical} level, we have learned that relativistic effects introduce an additional $m=1$ inner Lindblad resonance at which the pericentre precession rate of the test particle matches the pattern speed of the perturbation.  This has no Newtonian Keplerian analogue, but in quasi-Newtonian language one can think of it as being due to the steepening of the potential.  Indeed, any Newtonian potential with an ISCO will have this resonance.  We found, however, that the quasi-Newtonian calculation of the resonant strength, which is due to the tidal octupole, is suppressed by tens of percents due to gravitomagnetic corrections even at $r_0/M>100$.  In this sense the $m=1$ ILR is a relativistic beast.

At the {\em mathematical} level, our method of computation has revealed a connection between, on the one hand, angular momentum transfer via the Lindblad resonances; and on the other hand, the product of the gravitational wave signals emitted to infinity and into the hole by the perturber and the test particle (assuming the latter to be in an orbit of infinitesimal eccentricity).  This connection arose from general principles: (i) the conservation of energy and angular momentum when the contribution to both from gravitational waves is included; (ii) the fact that, aside from the $\ell=0$ and 1 modes that do not contribute to resonant transfer, the entire perturbed spacetime structure in the vacuum regions is determined by the radiation degrees of freedom, described for Type D spacetimes by $\psi_4$; and (iii) the ability to describe epicyclic motion of the test particle via Hamiltonian dynamics.  This was not expected when we began the calculation, and we are still lacking an intuitive explanation.

The relativistic corrections to the Lindblad resonance formulae -- particularly the existence of the new $m=1$ ILR and the strengthening of the $m\ge2$ ILRs -- may be important in binary black hole merger scenarios that involve an inner disc. This is especially true for the proposal of \citet{2009arXiv0906.0825C}, in which a bright electromagnetic counterpart is produced by resonant heating of this inner disc. A more full treatment of disc evolution including the new resonance as well as other Newtonian aspects of disc physics is beyond the scope of this paper; however, simple considerations suggest that this would be a fruitful exercise. \citet{2009arXiv0906.0825C} computed the inner disc evolution for a primary hole of mass $M=10^7M_\odot$ and mass ratio $q=0.1$, used Newtonian formulae for the torque, and treated the resonant torques as continuously distributed in radius (which may be appropriate for sufficiently small $|r_0-r_1|$). They find that the inner disc is truncated at $r_1<0.63r_0$ until $r_0\approx 20M$ (see Figs. 3 and 4 of \citealt{2009arXiv0906.0825C}); it is thus plausible that in a full treatment including the discrete nature of the Lindblad resonances, the strong $m=2$ ILR would truncate the disc. If this is the case, then even the weak $m=1$ ILR could be a significant contributor to resonant heating: while it is 3 orders of magnitude weaker than the $m=2$ ILR at $r_0=20$, if material in the $m=2$ ILR has been mostly cleared it is no longer obvious which resonance dominates the torque. This is especially true for retrograde configurations, where the $m=1$ ILR is enhanced. While the distribution of values of $a$ is presently quite uncertain, in the context of electromagnetic counterparts to a low-frequency gravitational wave detector such as the {\slshape Laser Interferometer Space Antenna} the value of $a$ for each event will in many cases be known to high precision \citep[e.g.][]{2006PhRvD..74l2001L}. Due to the weakness of the $m=1$ ILR, it may also be important to account for other weak resonances, e.g. inclination resonances in the case of a spinning primary; we have not computed the strengths of inclination resonances in this paper, but note that the techniques described here should be applicable to that problem.

\section*{Acknowledgments}

C.H. thanks Tanja Hinderer, Mike Kesden, and Dave Tsang for numerous helpful conversations.

C.H. is supported by the U.S. Department of Energy under contract DE-FG03-02-ER40701, the National Science Foundation under contract AST-0807337, and the Alfred P. Sloan Foundation.

\appendix

\section{Spheroidal harmonics}
\label{app:sph}

This appendix considers the solution to the angular eigenmode equation, Eq.~(\ref{eq:S}), for $S^{s,\chi}_{\ell,m}(\theta)$.

The most convenient way to solve Eq.~(\ref{eq:S}) is to write the eigenfunctions as linear combinations of the spin-weighted spherical harmonics
\citep{1966JMP.....7..863N, 1967JMP.....8.2155G}, as has been done in previous works \citep[e.g.][]{1973ApJ...185..649P, 2000PhRvD..61h4004H}
\begin{equation}
S^{\chi}_{\ell,m}(\theta) = \sum_{j=\max(|m|,|s|)}^\infty b_{j\ell m}(\chi) Y^s_{\ell,m}(\theta),
\end{equation}
where the coefficients $b_{j\ell m}(\chi)$ satisfy the eigenvalue equation \citep[][\S III{\slshape a}]{1973ApJ...185..649P},
\begin{equation}
{\mathbfss C}{\bmath b} = {\cal E}^s_{\ell,m} {\bmath b},
\end{equation}
where ${\bmath b}$ is a vector of length $j_{\rm max}-j_{\rm min}+1$ where $j_{\rm min}=\ell_{\rm min}$ and $j_{\rm max}$ is the highest angular momentum harmonic used in the finite basis set.  The matrix ${\mathbfss C}$ is real and symmetric, and is band-diagonal in the sense that $C_{jj'}=0$ if $|j-j'|>2$ \citep{1973ApJ...185..649P}.  In numerical computation, we truncate at $j_{\rm max}$, obtain the eigenvalues ${\cal E}^s_{\ell,m}$ and eigenvectors ${\bmath b}$ by Jacobi iteration, and compute the residual
\begin{equation}
e = \sqrt{\sum_{j>j_{\rm max}} ({\mathbfss C}{\bmath b})_{j}^2};
\end{equation}
$j_{\rm max}$ is increased until $e$ falls below some error threshold (usually $10^{-8}$) for all desired $\ell$.  The eigenvectors are normalized using $\sum_j b_{j\ell m}^2=2\pi$, which is equivalent to the usual normalization,
\begin{equation}
\int_0^\pi \sin\theta\,\left| S^{s,\chi}_{\ell,m}(\theta) \right|^2\,\rmd\theta = 1.
\end{equation}

The spin-weighted spherical harmonics are computed directly from the rotation matrices,
\begin{equation}
Y^s_{\ell,m}(\theta) = (-1)^m [{\mathbfss D}(\theta)]_{-s,m} =
 (-1)^m\left[ \exp (\rmi\theta {\mathbfss L}_2) \right]_{-s,m},
\end{equation}
where ${\mathbfss L}_2$ is the angular momentum operator around the 2-axis in the spin-$\ell$ representation of SO(3).\footnote{With the standard (Condon-Shortley) phases, $\rmi{\mathbfss L}_2$ is real and antisymmetric.}  The complex exponential is computed by a quadratic expansion for small $\theta$ ($\theta<10^{-8}$), and for larger values by repeated squaring of the rotation matrix ${\mathbfss D}(\theta)$ (each squaring doubles $\theta$).  For this process, we actually store ${\mathbfss D}(\theta)-{\mathbfss 1}$ where ${\mathbfss 1}$ is the $(2\ell+1)\times(2\ell+1)$ identity matrix; this is numerically preferable for small $\theta$ to avoid exponential amplification of rounding errors in the squaring process.  The squaring process is then
\begin{equation}
{\mathbfss D}(2\theta)-{\mathbfss 1} = 2[{\mathbfss D}(\theta)-{\mathbfss 1}] + [{\mathbfss D}(\theta)-{\mathbfss 1}]^2.
\end{equation}
This method is slow but is stable, simple to code, and does not suffer from underflow occurrences (common in many publicly available spherical harmonics routines even at modest $\ell$).  It also returns estimates of the $\theta$-derivatives with no extra effort since
\begin{equation}
\frac{\rmd {\mathbfss D}(\theta)}{\rmd\theta} = \rmi {\mathbfss L}_2 {\mathbfss D}(\theta).
\end{equation}

For evaluation of the source terms, we require $\opL_2^\dagger S$ and $\opL_1^\dagger\opL_2^\dagger S$.  Given $S$ and $\partial_\theta S$, it is easy to compute
\begin{equation}
\opL_2^\dagger S = \partial_\theta S +(- m\csc\theta + \chi \sin\theta + 2\cot\theta)S.
\label{eq:L2d}
\end{equation}
We further see that
\begin{eqnarray}
\opL_1^\dagger \opL_2^\dagger S \!\! &=& \!\! \partial_\theta^2 S + (-2m\csc\theta+2\chi\sin\theta+3\cot\theta)\partial_\theta S
\nonumber \\ && \!\!
+ (m^2\csc^2\theta +\chi^2\sin^2\theta - 2 - 2m\chi
\nonumber \\ && \;\;
 - 2m\csc\theta\cot\theta + 4\chi\cos\theta)S.
\end{eqnarray}
We may now use the angular Teukolsky equation for $S$, which is a second-order ODE that expresses $\partial_\theta^2S$ in terms of $S$, $\partial_\theta S$, and the eigenvalue ${\cal E}$.  Substituting out $\partial_\theta^2 S$, we find
\begin{eqnarray}
\opL_1^\dagger \opL_2^\dagger S \!\! &=& \!\!
2(-m\csc\theta+\chi\sin\theta+\cot\theta)\partial_\theta S
\nonumber \\ && \!\!
+ [-\chi^2\cos 2\theta - 2m\chi + 2m^2\csc^2\theta
\nonumber \\ && \;\; - 6m\csc\theta\cot\theta -2+4\csc^2\theta - {\cal E}]S,
\label{eq:L1L2d}
\end{eqnarray}
which is the form we use.

\section{Scattering matrix}
\label{app:wronskian}

Here we concern ourselves with the scattering matrix relating the four solutions of the radial Teukolsky equation,
\begin{equation}
\left( \begin{array}{c} {\cal R}_1(r) \\ {\cal R}_2(r) \end{array} \right)
=
\left( \begin{array}{cc} c_{13} & c_{14} \\ c_{23} & c_{24} \end{array} \right)
\left( \begin{array}{c} {\cal R}_3(r) \\ {\cal R}_4(r) \end{array} \right),
\label{eq:cmatrix}
\end{equation}
where the $c_{ab}$ are 4 complex coefficients that we wish to compute.  (We may also want the inverse matrix.)  Our goal here is the numerical computation of the $c_{ab}$ coefficients analytically from $\aleph$ and the parameters of the problem.

The Wronskian of any two solutions is $W_{ab} = {\cal R}_a{\cal R}'_b-{\cal R}_b{\cal R}'_a$ and is proportional to $\Delta$.  In particular, the asymptotic solutions give at the horizon gives
\begin{equation}
W_{12} = \left. 2(\rmi\varpi-\Gamma) \frac{\rmd r_\star}{\rmd r}\right|_{r_{{\rm h}+}} \Delta^2
= 2\beta\Delta,
\end{equation}
where
\begin{equation}
\beta \equiv 2\rmi Mr_{{\rm h}+}\omega - \rmi am - 2\sqrt{M^2-a^2}.
\end{equation}
The solutions at large radius give $W_{34}=-2\rmi\omega\Delta$.  We have also set $W_{31}=\aleph\Delta$.

The above Wronskian elements constrain the $c_{ab}$.  First, Eq.~(\ref{eq:cmatrix}) sets $W_{12}$ equal to $W_{34}$ times the determinant of the matrix of $c_{ab}$, so:
\begin{equation}
c_{13}c_{24}-c_{14}c_{23} = \rmi\frac{\beta}{\omega}.
\label{eq:cdet}
\end{equation}
Second, the definition of $\aleph$ implies that $\aleph\Delta = -c_{14}W_{34}$, so
\begin{equation}
c_{14} = -\rmi\frac{\aleph}{2\omega}.
\label{eq:c14}
\end{equation}

\cmnt{the power radiated to future null infinity is
\begin{equation}
P_{\rm out} = \omega^{-2}|b_3|^2,
\end{equation}
and that radiated into the hole is
\begin{equation}
P_{\rm down} = \alpha\omega^{-2}|b_1|^2.
\end{equation}
The power arriving from past null infinity can be obtained via reversal of $t$ and $\phi$, under which $(m,\omega)\leftrightarrow(-m,-\omega)$ and the $R_1$ and $R_2$ solutions are switched; the coefficient may be found using \citet[][Eq.~2.5]{1974ApJ...193..443T} and then the Teukolsky-Starobinsky identity \citep[][Eq.~3.22]{1974ApJ...193..443T})
\begin{equation}
b_3(m,\omega) \rightarrow \frac{16\omega^4}{C} b_4(-m,-\omega).
\end{equation}
This gives
\begin{equation}
P_{\rm in} = \frac{256\omega^6}{|C|^2}|b_4|^2
\end{equation}
\citep[][Eq. 4.12b]{1974ApJ...193..443T}.  We may also compute the power arriving from the past horizon by a similar method: under time reversal, and this time using the other Teukolsky-Starobinsky identity \citep[][Eq.~3.28]{1974ApJ...193..443T},
\begin{equation}
b_1(m,\omega) \rightarrow \frac{C^\ast}{(2Mr_{{\rm h}+})^4}\left[ \prod_{n=-1}^2 (2\rmi\varpi+n\Gamma)\right]^{-1} b_2(-m,-\omega).
\end{equation}
Thus
\begin{equation}
P_{\rm up} = \alpha'\omega^{-2}|b_2|^2,
\end{equation}
where
\begin{eqnarray}
\alpha' \!\! &=& \!\! \frac{\alpha |C|^2}{16(2Mr_{{\rm h}+})^8\varpi^2(4\varpi^2+\Gamma^2)^2(\varpi^2+\Gamma^2)}
\nonumber \\ &=& \!\!
\frac{2(\varpi^2+4\Gamma^2)\omega^3}{(Mr_{{\rm h}+})^3\varpi(4\varpi^2+\Gamma^2)^2}.
\end{eqnarray}
}

Further relations can be found from considering the conservation of energy.  For a general case with
\begin{equation}
{\cal R}(r) = b_1{\cal R}_1(r) + b_2{\cal R}_2(r) = b_3 {\cal R}_3(r) + b_4 {\cal R}_4(r),
\end{equation}
the conservation of energy \citep{1974ApJ...193..443T} then provides the relation
\begin{equation}
|b_3|^2 + \alpha|b_1|^2 = \frac{(2\omega)^8}{|C|^2}|b_4|^2 + \alpha_2|b_2|^2.
\label{eq:econs}
\end{equation}
Here the $b_2$ term denotes power emerging from the past horizon, whose value is not required here.
This relation may be evaluated for the case of ${\cal R}={\cal R}_1+\sigma{\cal R}_2$; equating terms on both sides proportional to $1$ and $\sigma$ (or $\sigma^\ast$) gives respectively
\begin{equation}
|c_{13}|^2 + \alpha = \frac{(2\omega)^6}{|C|^2}|\aleph|^2
\label{eq:cplex1}
\end{equation}
and
\begin{equation}
c_{23}^\ast c_{13} = \frac{(2\omega)^8}{|C|^2} c_{24}^\ast c_{14}.
\label{eq:cplex2}
\end{equation}

Equation~(\ref{eq:cplex2}) enables us to solve for $c_{23}$ in terms of the other coefficients;
substituting into the determinant relation, Eq.~(\ref{eq:cdet}), eliminates $c_{23}$ and generates a linear equation for $c_{24}$ in terms of $c_{13}$ and $c_{14}$:
\begin{equation}
\left[-\frac{(2\omega)^8c_{14}^\ast}{|C|^2c_{13}^\ast} c_{14} + c_{13} \right] c_{24} = \rmi\frac\beta\omega.
\end{equation}
Using Eq.~(\ref{eq:cplex1}) and substituting for $c_{14}$ (from Eq.~\ref{eq:c14}) simplifies this to
\begin{equation}
c_{24} = -\frac{\rmi \beta c_{13}^\ast}{\alpha\omega},
\label{eq:c24}
\end{equation}
and hence
\begin{equation}
c_{23} = -\frac{\rmi\beta}{\alpha\omega} \frac{(2\omega)^8}{|C|^2} c_{14}^\ast.
\label{eq:c23}
\end{equation}

The programme to compute the $c_{ab}$ is thus as follows:
\begin{itemize}
\item First obtain $c_{14}$ from Eq.~(\ref{eq:c14}) and the solution for $\aleph$ from Sec.~\ref{ss:inhomo}.
\item Next obtain $c_{13}$ by integrating the ${\cal R}_1$ solution along the real axis to large $r$, where the ${\cal R}_3$ solution becomes dominant.  By dividing by the asymptotic form for ${\cal R}_3$ (again keeping the first two coefficients in the expansion), obtain the coefficient of ${\cal R}_3$ in ${\cal R}_1$, i.e.$c_{13}$.
\item Evaluate $\beta$ and then use Eqs.~(\ref{eq:c24}) and (\ref{eq:c23}) to obtain $c_{24}$ and $c_{23}$.
\end{itemize}

The inverse transformation coefficients $c_{31}$, $c_{32}$, $c_{41}$, and $c_{42}$ can be obtained in accordance with
\begin{equation}
\left(\begin{array}{cc} c_{31} & c_{32} \\ c_{41} & c_{42} \end{array}\right)
= \frac\omega{\rmi\beta}
\left(\begin{array}{cc} c_{24} & -c_{14} \\ -c_{23} & c_{13} \end{array}\right);
\end{equation}
we note that the substitution of the formula for the determinant in the denominator is required if this relation is used for numerical computation because of the very large correlation coefficient of the matrix, i.e. for some practical cases we have $c_{13}c_{24}\approx c_{14}c_{23}$.  However, for formulas involving $c_{31}$ it is more convenient to combine this with Eq.~(\ref{eq:c24}) to obtain
\begin{equation}
c_{31} = -\frac{c_{13}^\ast}{\alpha}.
\label{eq:c31}
\end{equation}

\section{Radial modes at low frequency}
\label{app:R0}

This appendix describes the radial modes in the nonrelativistic regime, i.e. where $\omega\ll M^{-1}$ and $M\ll r\ll \omega^{-1}$.  This is the regime relevant for Newtonian Keplerian discs (Section~\ref{sec:Kepler}).  The angular modes simply reduce to spin-weighted spherical harmonics with separation constant ${\cal E}=\ell(\ell+1)$.

There are infinite (logarithmically divergent in $r$ or $r-r_{{\rm h}+}$) phase errors in our approximations here; this does not concern us since the absolute phases of ${\cal R}_1$ at the horizon or ${\cal R}_3$ at infinity cancel out of the computation.

The solution of the radial modes in terms of $_1F_1$ functions is described in greater generality by \citet{1996PThPh..95.1079M}; see also the review by \citet[][\S4]{2003LRR.....6....6S}.  We sketch here a simplified derivation for the specialized case of small $\omega$, which does not require a ``renormalized angular momentum parameter'' and has much shorter expressions.

\subsection{The ${\cal R}_1$ solution}

The ${\cal R}_1$ solution (no radiation emerging from the past horizon) in this regime can be constructed by taking $\omega\rightarrow 0$.  With this simplification, the radial Teukolsky equation can be reduced to a hypergeometric equation \citep{1996PThPh..95.1079M}.  The solution is
\begin{equation}
{\cal R} \propto (-x)^{2-\rmi\tau/2} (1-x)^{-\rmi\tau/2}\;_2F_1(\ell+1-\rmi\tau,-\ell-\rmi\tau;3-\rmi\tau;x),
\end{equation}
where
\begin{equation}
x = \frac{r_{{\rm h}+}-r}{r_{{\rm h}+}-r_{{\rm h}-}} = -\frac{r-r_{{\rm h}+}}{2\sqrt{M^2-a^2}}
\end{equation}
and $\tau = -am/\sqrt{M^2-a^2}$.  Outside the horizon we have $x<0$, and we take the branch $\arg(1-x)=\arg(-x)=0$ of the fractional powers.


The normalization of ${\cal R}_1$ can be obtained by taking the limit as $r\rightarrow r_{{\rm h}+}^+$ ($-x\rightarrow 0^+$).  This gives
\begin{equation}
{\cal R}_1\rightarrow \Delta^2 \rme^{\rmi m\Omega_{\rm H} r_\star}
\approx 16(M^2-a^2)^2 \rme^{\rmi m\varphi_0}
(-x)^{2-\rmi\tau/2},
\end{equation}
where $\varphi_0 = \frac12 a_\star + \Omega_{\rm H}M\ln(1-a_\star^2)$ and we have substituted for $\Omega_{\rm H}$ in order to simplify the exponent of $-x$.  We thus see that
\begin{eqnarray}
\lim_{\omega\rightarrow 0}
{\cal R}_1(r) \!\! &=& \!\! 16(M^2-a^2)^2 \rme^{\rmi m\varphi_0} (-x)^{2-\rmi\tau/2} (1-x)^{-\rmi\tau/2}
\nonumber \\ && \!\!\times
\;_2F_1(\ell+1-\rmi\tau,-\ell-\rmi\tau;3-\rmi\tau;x).
\end{eqnarray}
The series can be made finite using the linear transformation formula \citep[][Eq.~15.3.3]{1972hmfw.book.....A}:
\begin{eqnarray}
\lim_{\omega\rightarrow 0}
{\cal R}_1(r) \!\! &=& \!\! 16(M^2-a^2)^2 \rme^{\rmi m\varphi_0} (-x)^{2-\rmi\tau/2} (1-x)^{2+\rmi\tau/2}
\nonumber \\ && \!\!\times
\;_2F_1(2-\ell, 3+\ell; 3-\rmi\tau; x).
\end{eqnarray}
To reach the Keplerian regime, we must follow this to the regime where $-x\gg 1$.  Taking the highest-order ($r^{\ell-2}$) term in the series, we find
\begin{eqnarray}
{\cal R}_1(r) \!\!&\approx& \!\! -16(M^2-a^2)^2 \rme^{\rmi m\varphi_0}
\frac{(2\ell)!}{(\ell+2)!}\frac{\Gamma(3-\rmi\tau)}{\Gamma(\ell+1-\rmi\tau)}
\nonumber \\ && \!\!\times
\left( \frac{r}{2\sqrt{M^2-a^2}} \right)^{\ell+2}.
\end{eqnarray}
For our purposes, this may be written as
\begin{equation}
{\cal R}_1(r) \rightarrow k_1 r^{\ell+2},
\label{eq:R1K}
\end{equation}
where using the recursion relation for the $\Gamma$ function\footnote{The product is empty for $\ell=2$, in which case it is understood to evaluate to unity.},
\begin{equation}
|k_1| = [2(M^2-a^2)]^{(2-\ell)/2} \frac{(2\ell)!}{(\ell+2)!} \left[\prod_{n=3}^\ell (n^2+\tau^2)\right]^{-1/2}.
\end{equation}
We will not require the {\em phase} of $k_1$; indeed, the phase is meaningless at the level of approximation here because in taking $\omega\rightarrow 0$ we introduce a phase error of $\sim \omega|r_\star|$, which diverges as one approaches the horizon.

\subsection{The ${\cal R}_3$ solution}

We are now interested in the solutions that asymptote to a purely outgoing wave at $r\rightarrow\infty$.  In this case, we keep $\omega$ but approximate $M,a\rightarrow 0$.  \citet{1996PThPh..95.1079M} also provides a solution in this case in terms of a confluent hypergeometric function.  They find ${\cal R}\propto rf$, where $z=\omega r$ and $f(z)$ satisfies the relation
\begin{equation}
z^2 \frac{\rmd^2f}{\rmd z^2} + [z^2-4\rmi z-\ell(\ell+1)]f=0.
\label{eq:cc}
\end{equation}
As is well-known, this equation reduces to a $_1F_1$-type series upon the substitution $f(z) = \rme^{\pm\rmi z}g(z)$.  Four solutions may be obtained this way, depending on the chosen leading power of $z$:
\begin{eqnarray}
{\cal R}_{\rm A}(r) \!\! &=& \!\! r^{\ell+2}\rme^{-\rmi\omega r} \;_1F_1(\ell+3;2\ell+2;2\rmi\omega r),
\nonumber \\
{\cal R}_{\rm B}(r) \!\! &=& \!\! r^{1-\ell}\rme^{-\rmi\omega r} \;_1F_1(2-\ell;-2\ell;2\rmi\omega r),
\nonumber \\
{\cal R}_{\rm C}(r) \!\! &=& \!\! r^{\ell+2}\rme^{\rmi\omega r}\;_1F_1(\ell-1;2\ell+2;-2\rmi\omega r), {\rm ~~and}
\nonumber \\
{\cal R}_{\rm D}(r) \!\! &=& \!\! r^{1-\ell}\rme^{\rmi\omega r}\;_1F_1(-2-\ell;-2\ell;-2\rmi\omega r).
\end{eqnarray}
Of these, ${\cal R}_{\rm B}$ and ${\cal R}_{\rm D}$ have the advantage of having truncating (polynomial) $_1F_1$ series; due to the nature of their oscillating parts they are manifestly linearly independent and provide a complete basis.  The highest power in $r$ shows that ${\cal R}_{\rm B}(z)$ yields the ${\cal R}_4$ solution and ${\cal R}_{\rm D}(z)$ yields the ${\cal R}_3$ solution.  The normalization is easily obtained from the highest term:
\begin{equation}
{\cal R}_3(r) = \rmi^{\ell-2} \frac{(2\ell)!\,r^{1-\ell}\rme^{\rmi\omega r}}{(\ell-2)!\,(2\omega)^{\ell+2}}
\;_1F_1(-2-\ell;-2\ell;-2\rmi\omega r).
\end{equation}
This is only valid in the limiting case where $M\rightarrow 0$; finite mass introduces a logarithmically divergent phase error due to the long-range nature of the background metric perturbation (the asymptotic expansion of $\rmd r_\star/\rmd r-1$ begins with the order $r^{-1}$ term).

For the Newtonian Keplerian problem, we require the near-field solution $r\ll\omega^{-1}$, where
\begin{equation}
{\cal R}_3(r) = k_3r^{1-\ell}
\label{eq:R3K}
\end{equation}
with
\begin{equation}
k_3 = \rmi^{\ell-2} \frac{(2\ell)!}{(\ell-2)!\,(2\omega)^{\ell+2}}.
\end{equation}

A similar result allows us to normalize ${\cal R}_4$: in the near-field zone, ${\cal R}_4(r) =k_4r^{1-\ell}$ with
\begin{equation}
k_4 = \rmi^{-\ell-2} \frac{(2\ell)!}{(\ell+2)!}(2\omega)^{2-\ell}.
\end{equation}

\subsection{Wronskians and scattering coefficients}

The Wronksian of the ${\cal R}_1$ and ${\cal R}_3$ solutions is easily evaluated in the Keplerian range of radii.  It leads to
$\aleph = (2\ell+1)k_1k_3$,
hence
\begin{equation}
\aleph = \rmi^{\ell-2} \frac{(2\ell+1)!}{(\ell-2)!\,(2\omega)^{\ell+2}}k_1.
\end{equation}

Finally, for resonant amplitude problems we will require $c_{13}$ from ${\cal R}_1=c_{13}{\cal R}_3+c_{14}{\cal R}_4$.  We see that in the near-field region $M\ll r\ll \omega^{-1}$, ${\cal R}_1$ is dominated by the growing-outward ($r^{\ell+2}$) solution while ${\cal R}_3$ and ${\cal R}_4$ are both dominated by the growing-inward ($r^{1-\ell}$) solution.  Therefore the ratio $c_{13}:c_{14}$ can be obtained by forcing the leading terms inward (i.e. coefficients of $\propto r^{1-\ell}$) to cancel.  This is
\begin{equation}
c_{13} = -\frac{k_4}{k_3}c_{14} = \rmi \frac{k_4\aleph}{2\omega k_3} = \rmi^{-\ell+1} \frac{(2\ell+1)!}{(\ell+2)!} (2\omega)^{1-\ell} k_1.
\end{equation}


\begin{thebibliography}{}

\bibitem[\protect\citeauthoryear{{Abramowitz} \& {Stegun}}{1972}]{1972hmfw.book.....A}
{Abramowitz} M., {Stegun} I., 1972, Handbook of Mathematical Functions, Dover, New York, NY

\bibitem[\protect\citeauthoryear{{Arnold}}{1978}]{1978mmcm.book.....A}
{Arnold} V., 1978, Mathematical Methods of Classical Mechanics, Springer, New York, NY

\bibitem[\protect\citeauthoryear{{Boyer} \& {Lindquist}}{1967}]{1967JMP.....8..265B}
{Boyer} R., {Lindquist} R., 1967, J. Math. Phys., 8, 265

\bibitem[\protect\citeauthoryear{{Chandrasekhar}}{1992}]{1992mtbh.book.....C}
{Chandrasekhar} S., 1992, The Mathematical Theory of Black Holes, Oxford University Press, New York, NY

\bibitem[\protect\citeauthoryear{{Chang} et~al.}{2010}]{2009arXiv0906.0825C}
{Chang} P., {Strubbe} L., {Menou} K., {Quataert} E., 2010, MNRAS, 407, 2007

\bibitem[\protect\citeauthoryear{{Chrzanowski}}{1975}]{1975PhRvD..11.2042C}
{Chrzanowski} P., 1975, Phys. Rev. D, 11, 2042

\bibitem[\protect\citeauthoryear{{Detweiler}}{1978}]{1978ApJ...225..687D}
{Detweiler} S., 1978, ApJ, 225, 687

\bibitem[\protect\citeauthoryear{{Drasco} \& {Hughes}}{2006}]{2006PhRvD..73b4027D}
{Drasco} S., {Hughes} S., 2006, Phys. Rev. D, 73, 024027

\bibitem[\protect\citeauthoryear{{Drasco} et~al.}{2005}]{2005CQGra..22S.801D}
{Drasco} S., {Flanagan} \'E., {Hughes} S., 2005, Cl. Quant. Grav., 22, S801

\bibitem[\protect\citeauthoryear{{Flanagan} \& {Hinderer}}{2010}]{2010arXiv1009.4923F}
{Flanagan} \'E., {Hinderer} T., 2010, preprint, arXiv:1009.4923


\bibitem[\protect\citeauthoryear{{Goldberg} et~al.}{1967}]{1967JMP.....8.2155G}
{Goldberg} J., {Macfarlane} A., {Newman} E., {Rohrlich} F., {Sudarshan} E., 1967, J. Math. Phys., 8, 2155

\bibitem[\protect\citeauthoryear{{Goldreich} \& {Tremaine}}{1978}]{1978ApJ...222..850G}
{Goldreich} P., {Tremaine} S., 1978, ApJ, 222, 850

\bibitem[\protect\citeauthoryear{{Goldreich} \& {Tremaine}}{1979}]{1979ApJ...233..857G}
{Goldreich} P., {Tremaine} S., 1979, ApJ, 233, 857

\bibitem[\protect\citeauthoryear{{Goldreich} \& {Tremaine}}{1980}]{1980ApJ...241..425G}
{Goldreich} P., {Tremaine} S., 1980, ApJ, 241, 425

\bibitem[\protect\citeauthoryear{{Goldstein} et~al.}{2002}]{2002clme.book.....G}
{Goldstein} H., {Poole} C., {Safko} J., 2002, Classical Mechanics, 3rd ed., Addison-Wesley, San Francisco, USA

\bibitem[\protect\citeauthoryear{{Hinderer} \& {Flanagan}}{2008}]{2008PhRvD..78f4028H}
{Hinderer} T., {Flanagan} \'E., 2008, Phys. Rev. D, 78, 064028

\bibitem[\protect\citeauthoryear{{Hughes}}{2000}]{2000PhRvD..61h4004H}
{Hughes} S., 2000, Phys. Rev. D, 61, 084004

\bibitem[\protect\citeauthoryear{{Hughes} et~al.}{2005}]{2005PhRvL..94v1101H}
{Hughes} S., {Drasco} S., {Flanagan} \'E., {Franklin} J., 2005, Phys. Rev. Lett., 04, 221101


\bibitem[\protect\citeauthoryear{{Kennefick}}{1998}]{1998PhRvD..58f4012K}
{Kennefick} D., 1998, Phys. Rev. D, 58, 064012

\bibitem[\protect\citeauthoryear{{Lang} \& {Hughes}}{2006}]{2006PhRvD..74l2001L}
{Lang} R., {Hughes} S., 2006, Phys. Rev. D, 74, 122001

\bibitem[\protect\citeauthoryear{{Lin} \& {Papaloizou}}{1979}]{1979MNRAS.186..799L}
{Lin} D., {Papaploizou} J., 1979, MNRAS, 186, 799

\bibitem[\protect\citeauthoryear{{Lynden-Bell} \& {Kalnajs}}{1972}]{1972MNRAS.157....1L}
{Lynden-Bell} D., {Kalnajs} A., 1972, MNRAS, 157, 1

\bibitem[\protect\citeauthoryear{{Mano} et~al.}{1996}]{1996PThPh..95.1079M}
{Mano} S., {Suzuki} H., {Takasugi} E., 1996, Prog. Theor. Phys., 95, 1079

\bibitem[\protect\citeauthoryear{{Mino} et~al.}{1997}]{1997PThPS.128....1M}
{Mino} Y., {Sasaki} M., {Shibata} M., {Tagoshi} H., {Tanaka} T., 1997, Prog. Theor. Phys. Supp., 128, 1

\bibitem[\protect\citeauthoryear{{Mino}}{2003}]{2003PhRvD..67h4027M}
{Mino} Y., 2003, Phys. Rev. D, 67, 084027

\bibitem[\protect\citeauthoryear{{Misner} et~al.}{1973}]{1973grav.book.....M}
{Misner} C., {Thorne} K., {Wheeler} J., 1973, Gravitation, W. H. Freeman and Co., San Francisco, California, USA

\bibitem[\protect\citeauthoryear{{Murray} \& {Dermott}}{2000}]{2000ssd..book.....M}
{Murray} C., {Dermott} S., 2000, Solar System Dynamics, Cambridge University Press, Cambridge, UK

\bibitem[\protect\citeauthoryear{{Newman} \& {Penrose}}{1966}]{1966JMP.....7..863N}
{Newman} E., {Penrose} R., 1966, J. Math. Phys., 7, 863

\bibitem[\protect\citeauthoryear{{Okazaki} et~al.}{1987}]{1987PASJ...39..457O}
{Okazaki} A., {Kato} S., {Fukue} J., 1987, PASJ, 39, 457

\bibitem[\protect\citeauthoryear{{Ori}}{2003}]{2003PhRvD..67l4010O}
{Ori} A., 2003, Phys. Rev. D, 67, 124010

\bibitem[\protect\citeauthoryear{{Porco} et~al.}{1984}]{1984Icar...60....1P}
{Porco} C., {Nicholson} P., {Borderies} N., {Danielson} G., {Goldreich} P., {Holberg} J., {Lane} A., 1984, Icarus, 60, 1

\bibitem[\protect\citeauthoryear{{Press} \& {Teukolsky}}{1973}]{1973ApJ...185..649P}
{Press} W., {Teukolsky} S., 1973, ApJ, 185, 649

\bibitem[\protect\citeauthoryear{{Press} et~al.}{1992}]{1992nrca.book.....P}
{Press} W., {Teukolsky} S., {Vetterling} W., {Flannery} B., 1992, Numerical recipes in C. The art of scientific computing, Cambridge University Press, Cambridge, UK

\bibitem[\protect\citeauthoryear{{Sasaki} \& {Nakamura}}{1982a}]{1982PhLA...89...68S}
{Sasaki} M., {Nakamura} T., 1982a, Phys. Lett. A, 89, 68

\bibitem[\protect\citeauthoryear{{Sasaki} \& {Nakamura}}{1982b}]{1982PThPh..67.1788S}
{Sasaki} M., {Nakamura} T., 1982b, Prog. Theor. Phys., 67, 1788

\bibitem[\protect\citeauthoryear{{Sasaki} \& {Tagoshi}}{2003}]{2003LRR.....6....6S}
{Sasaki} M., {Tagoshi} H., 2003, Living Reviews in Relativity, 6, 6

\bibitem[\protect\citeauthoryear{{Schmidt}}{2002}]{2002CQGra..19.2743S}
{Schmidt} W., 2002, Cl. Quant. Grav., 19, 2743

\bibitem[\protect\citeauthoryear{{Shibata}}{1993}]{1993PhRvD..48..663S}
{Shibata} M., 1993, Phys. Rev. D, 48, 663

\bibitem[\protect\citeauthoryear{{Shibata}}{1994}]{1994PhRvD..50.6297S}
{Shibata} M., 1994, Phys. Rev. D, 50, 6297

\bibitem[\protect\citeauthoryear{{Teukolsky}}{1973}]{1973ApJ...185..635T}
{Teukolsky} S., 1973, ApJ, 185, 635

\bibitem[\protect\citeauthoryear{{Teukolsky} \& {Press}}{1974}]{1974ApJ...193..443T}
{Teukolsky} S., {Press} W., 1974, ApJ, 193, 443

\bibitem[\protect\citeauthoryear{{Wald}}{1978}]{1978PhRvL..41..203W}
{Wald} R., 1978, Phys. Rev. Lett., 41, 203

\bibitem[\protect\citeauthoryear{{Wald}}{1984}]{1984ucp..book.....W}
{Wald} R., 1984, General Relativity, University of Chicago Press, Chicago, IL


\end{thebibliography}
\end{document}